\documentclass[prb,aps,twocolumn,superscriptaddress,showpacs]{revtex4-1}
\usepackage{amsmath,amsfonts,amssymb}
\usepackage{braket} 
\usepackage{ulem} 
\usepackage{framed} 
\usepackage[pdftex]{graphicx}
\usepackage{natbib}
\usepackage{bbm}    
\usepackage{color}
\usepackage{multirow} 

\newcommand{\bs}[1]{\boldsymbol{#1}}
\newcommand{\abs}[1]{\left|{#1}\right|}
\newcommand{\sign}{\,\text{sign}}

\begin{document}

\title{Emergent topological properties in interacting one-dimensional systems with spin-orbit coupling}

\author{Nikolaos Kainaris}
\affiliation{Institut f\"ur Nanotechnologie, Karlsruhe Institute of Technology, 76021 Karlsruhe, Germany}
\affiliation{Institut f\"ur Theorie der Kondensierten Materie, Karlsruher Institut f\"ur
  Technologie, 76128 Karlsruhe, Germany}
  
\author{Sam T. Carr} 
\affiliation{School of Physical Sciences, University of Kent, Canterbury CT2  7NH, United Kingdom}

\date{\today}

\begin{abstract}
   We present analysis of a single channel interacting quantum wire problem in the presence of spin-orbit interaction. The spin-orbit coupling breaks the spin-rotational
   symmetry from SU(2) to U(1) and breaks inversion symmetry. The low-energy theory is then a two band model with a difference of Fermi velocities $\delta v$.
   Using bosonization and a two-loop renormalization group procedure we show that electron-electron interactions can open a gap in the spin sector of the theory when the interaction strength $U$ is smaller than $\delta v$ in appropriate units. For repulsive interactions, the resulting strong coupling phase is of the spin-density-wave type. We show that this phase has peculiar emergent topological properties. The gapped spin sector behaves as a topological insulator, with zero-energy edge modes with fractional spin.  On the other hand, the charge sector remains critical, meaning the entire system is metallic.  However, this bulk electron liquid as a whole exhibits properties commonly associated with the one-dimensional edge states of two-dimensional spin-Hall insulators, in particular, the conduction of $2e^2/h$ is robust against nonmagnetic impurities.
\end{abstract}

\pacs{}

\maketitle

\section{Introduction}
\label{sec:introduction}

The discovery of the quantum spin Hall insulator \cite{Kane_Mele_2005a,Kane_Mele_2005b,Bernevig_2006, Koenig_2007} sparked the realization that gapped phases of matter with identical bulk spectra are not all equivalent. Indeed, if an insulator can be characterized by a nonzero topological invariant, it hosts gapless modes at its edge which are robust to perturbations that respect the anti-unitary symmetries of the system. This topological protection crucially depends on a finite gap in the bulk.
While the topological classification of non-interacting systems is well established,\cite{Schnyder_2010} the role of strong interactions is still a matter of ongoing research.

A prime example of a strongly correlated system are electrons in a quantum wire which are a good realization of a Tomonaga-Luttinger liquid (LL).\cite{GNT_book,Giamarchi_book}   The distinctive feature of this state of matter is that the elementary excitations have no relation to free electrons, but rather are described by collective plasmon modes. Additionally, these plasmon modes carry spin and charge independently, a phenomenon known as spin-charge separation. 
As a consequence of the collective nature of the elementary excitations, even weak interactions between the electrons have profound consequences for the quantum state of the system. If interactions become strong, they can lead to a strong coupling regime where spectral gaps are generated dynamically without spontaneous breaking of any continuous symmetry. 

A particularly fascinating example of such a dynamically generated state is the Luther-Emery liquid\cite{Luther_Emery_1974}, in which the charge sector remains critical (gapless), however, the spin degrees of freedom acquire a gap.  This can quite naturally occur for attractive interactions, in which case the pairs of spins form singlets and the system exhibits many properties of superconductivity.\cite{GNT_book,Giamarchi_book,Seidel_Lee_2005}   If the interactions are repulsive, however, the spin-gap may only form if SU(2) spin-rotational symmetry is somehow broken in the system.\cite{GNT_book,Giamarchi_book}  While this case has the same thermodynamic spectrum as the spin gap that appears for attractive interactions, the states are rather different, with the dominant correlations in the repulsive case being of the spin-density- wave (SDW) type.

In this work, we will argue that the spin-gapped system with \textit{repulsive interactions} is topologically nontrivial. This topological state is peculiar in the sense that (i) it emerges as a strong coupling phase of the original model and (ii) even while one sector of the theory is gapped the other remains gapless so that the overall electron liquid is not in a gapped phase. Nonetheless,the system exhibits properties usually associated with topological insulators. First, zero-energy edge modes with fractional spin emerge at the boundary of a finite system. We note that this property has recently been predicted in a related one-dimensional model with spin-triplet pairing in Ref.~\onlinecite{Keselman_2015}, while various topological properties of gapless states have been discussed more generally in Refs.~\onlinecite{Baum_2015} and  \onlinecite{Hetenyi_2013}. Second, since electrons carry both charge and spin the gap in the spin sector affects the whole electron liquid and leads to unusual transport properties. In particular, we find that the bulk of the wire is robust against nonmagnetic impurities as long as interactions are not too strong [for $K_c>3/4$]. We note that these transport properties are inherent to the SDW phase and have also been discussed in Ref. \onlinecite{Sun_2007} and \onlinecite{Gangadharaiah_2008}. Throughout the paper we will compare the transport properties of the two realizations in more detail. 

The study of strongly interacting one-dimensional systems is by no means academic. Over the last several years there has been remarkable progress in nanotechnology which has led to an explosive growth of experimental work on low-dimensional systems. Single and multichannel one-dimensional conductors or quantum wires can now be manufactured in a controlled fashion.\cite{Auslaender_2002,Auslaender_2005,Venkataraman_2006,Slot_2004,Lu_Lieber_2006, Shen_2009} These systems provide a fertile ground for laboratory experiments of strongly interacting systems. 

However, as interactions in quantum wires are naturally repulsive, the SDW phase can only be realized if the spin SU(2) symmetry in the systems is broken. In a realistic setup, this is naturally achieved by the presence of spin-orbital interaction. 
In experiment, ballistic quantum wires are created in a two-dimensional electron gas (2DEG) by cleaved edge overgrowth. They are therefore naturally subject to spin-orbit coupling (SOC) which breaks the SU(2) symmetry. 
The SOC arises due to the asymmetry associated with the potential that constricts electrons to the two-dimensional plane, the so-called Rashba SOC.\cite{Rashba_1984} The asymmetry and therefore the Rashba SOC can be further controlled by applying external gate voltage.\cite{Nitta_1997,Engels_1997,Grundler_2000} In addition to the noted asymmetry due to the confining potential (which include the quantum-well potential that confines the electrons to the 2D layer as well as the in-plane potential that forms the quantum wire \cite{Moroz_Barnes_1999}) spin-orbit interaction is inherent to semiconductors of zinc-blende or wurtzite structure lacking a center of inversion. This leads to the so-called Dresselhaus SOC.\cite{Dresselhaus_1955} In this paper, we consider the situation where the Rashba term is tuned to be much stronger than the Dresselhaus term. This limit can be achieved experimentally by applying a sufficiently strong backgate voltage.

The effect of SOC on interacting one-dimensional systems has been discussed extensively in the literature.\cite{Sun_2007,Gangadharaiah_2008,Moroz_Barnes_1999,Moroz_2000_PRB,Moroz_2000_PRL,
Haeusler_2001,Iucci_2003,Governale_2004,Gritsev_2005,Schulz_2009,Goth_2014} However, so far it was believed that the SDW state can only be realized with the help of an additionally applied magnetic field which explicitly breaks the time-reversal symmetry of the system\cite{Sun_2007,Gangadharaiah_2008} or by fine tuning a modulated Rashba SOC.\cite{Malard_2011,Japaridze_2014}
  
In contrast, we show in this work that the SDW state can be realized in quantum wires with strong spin-orbital interaction even without additional perturbations such as magnetic fields as long as the spin-orbit coupling leads to different velocities\cite{Moroz_Barnes_1999,Moroz_2000_PRB,Moroz_2000_PRL} at the Fermi points of the low-energy bands. In this case, we find the criterion for the gap opening that the dimensionless velocity difference $\overline{\delta v}$ should be larger than the dimensionless Hubbard $U$ interaction. This constitutes one of the main results of our work;
throughout the paper we will critically contrast our results with previous work on spin-orbit coupling in quantum wires in order to explain how we obtained a different answer.  Having shown that the SDW may be realized, we will then discuss the topological properties of this state.  In particular, we will show that zero-energy edge modes emerge at the boundary of a finite wire, and electrical conduction in this state is insensitive to impurities as long as time-reversal symmetry remains unbroken.  This constitutes the second main result of this paper.

The paper is structured as follows. In Sec.~\ref{sec:Tight binding model for one dimensional electrons in the presence of spin-orbit-coupling}, we first use a specific tight binding model to study the effect of spin-orbit interaction on the spectrum of non-interacting fermions. After having established some basic understanding, we summarize results of previous studies and contrast them to this work. In Sec.~\ref{sec:Interaction effects}, we use the results from the noninteracting case to formulate an effective low-energy theory in the presence of interactions. We proceed to bosonize this model which takes interactions into account exactly and discuss the influence of SOC by deriving the renormalization group (RG) flow in Sec.~\ref{subsec:Renormalization group analysis}. We find that the RG flows to strong coupling in a certain parameter regime which leads to the opening of a spin gap. The nature of the strong coupling fixed point is discussed in Sec. \ref{subsec:The nature of the strong coupling phase} and we study the topological properties as well as the effect of disorder in Sec.~\ref{sec:Properties of the spin-density wave phase}. A summary of the present work is presented in Sec.~\ref{sec:Summary}, which is followed by two appendixes. In Appendix~\ref{App:Bosonization conventions}, we present the bosonization and refermionization conventions used throughout the work, while Appendix~\ref{App:Details of the momentum space RG} presents details of the renormalization group procedure.
Throughout the paper, we use units where $\hbar =1$. 

\section{Tight binding model for one dimensional electrons in the presence of spin-orbit-coupling}
\label{sec:Tight binding model for one dimensional electrons in the presence of spin-orbit-coupling}

We consider spinful fermions confined to one spatial dimension and subject to a Rashba spin-orbit-interaction at incommensurate filling. The Hamiltonian of the model is
\begin{align}
   H = H_0 + H_{\text{SO}}+ H_{\text{int}} \, .\label{latticemodel}
\end{align}
The bare hopping Hamiltonian is given by
\begin{align}
\begin{split} 
   H_0 =& -t \sum_{j,\sigma} c_{j,\sigma}^{\dagger} c^{}_{j+1,\sigma}+ \text{H.c.} \\
                  & -t' \sum_{j,\sigma} c_{j,\sigma}^{\dagger} c^{}_{j+2,\sigma} +\text{H.c.} - \mu N \; \\
                 =& \; - \sum_{k,\sigma} \left[ 2 t \cos(k) + 2 t' \cos(2 k) +\mu  \right] c_{k,\sigma}^{\dagger} c^{}_{k,\sigma}\, .\label{H0}
\end{split}                 
\end{align}
Here $c_{j,\sigma}$ destroys an electron with spin $\sigma= \uparrow,\downarrow$ at site $j$. The hopping amplitude is denoted by $t$ for nearest neighbor and $t'$ for next-nearest-neighbor hopping. We use dimensions where the lattice spacing $a_0=1$ and we assume periodic boundary conditions.

\noindent
The Rashba-type SOC reads as
\begin{align}
\begin{split} 
   H_{\text{SO}} =& -i \alpha \sum_{j,\sigma,\sigma'} c_{j,\sigma}^{\dagger} \sigma_{\sigma,\sigma'}^z c_{j+1,\sigma'}
                             + \text{H.c.} \\
                           =& \; 2 \alpha \sum_{k} \sin(k) c_{k,\sigma}^{\dagger} \sigma_{\sigma,\sigma'}^z c_{k,\sigma'}^{} \; .     
                            \label{HSO}
\end{split}                            
\end{align} 
The Rashba SOC with coupling strength $\alpha$ breaks the SU(2) spin-rotational symmetry down to U(1) but preserves time-reversal symmetry. Here $\sigma^{i}$, with $i \in \lbrace x,y,z \rbrace$, denotes the set of Pauli matrices in spin space. Finally, the Hubbard interaction is given by
\begin{align}
   H_{\text{int}} = U \sum_j n_{j,\uparrow} n_{j,\downarrow} \, . \label{HHubbard}
\end{align}
with the coupling constant $U$ and the electron density operator $n_{j,\sigma} = c_{j,\sigma}^{\dagger} c_{j,\sigma}^{}$. 

Throughout this work, we assume the hopping amplitudes as positive, $t,t' >0$ and repulsive interactions, $U>0$. 

\subsection{Effects of spin orbit coupling on the spectrum of noninteracting electrons}
\label{subsec:Effects of spin orbit coupling on the spectrum of noninteracting electrons}
First, we discuss the effect of SOC on the spectrum of noninteracting fermions without next-nearest-neighbor coupling i.e we set $t'=0$.
The non-interacting part of the Hamiltonian~(\ref{latticemodel}) then reads as
\begin{align}
   H_0 = \sum_{k,\sigma} c_{k,\sigma}^{\dagger} \left[ -2 t \cos(k) +2 \alpha \sigma \sin(k)\right]  c_{k,\sigma}^{} \, \label{noninteractinghamiltonian},
\end{align}  
where $\sigma = \pm 1$ are the eigenvalues of $\sigma^z$. Using the harmonic addition theorem, this can be recast into the form
\begin{align}
  H_0 =&  -2 \tilde{t} \sum_{k,\sigma} \cos(k - \sigma q_0 )c_{k,\sigma}^{\dagger} c_{k,\sigma} \,.
\end{align}
Here, $\tilde{t} = \sqrt{t^2+\alpha^2}$ is the renormalized hopping amplitude and $q_0 = \arctan(\alpha / t)$.

We conclude that in the absence of next-nearest-neighbor hopping SOC renormalizes the hopping amplitude and shifts the spectrum by a constant momentum $\pm q_0$, for spin up or down respectively.  
However, this shift can always be removed by a spin dependent gauge transformation and therefore has no observable effect on the thermodynamic properties of the system [they depend only on the spectrum].

This statement can be made explicit by considering the Hamiltonian in Eq.~(\ref{noninteractinghamiltonian}) back in real space: 
\begin{align}
   \tilde{H}_0 =&  -2 \tilde{t} \sum_{j,\sigma} e^{i q_0 j \sigma} c_{j,\sigma}^{\dagger} c_{j+1,\sigma}^{}
                         =  -2 \tilde{t} \sum_{j,\sigma} d_{j,\sigma}^{\dagger} d_{j+1,\sigma}^{} \, ,
\end{align}
where we defined the fermion operator $d_{j,\sigma} = e^{i q_0 j \sigma} c_{j,\sigma}$. The model in the presence of SOC is therefore unitarily equivalent to a model without SOC but with renormalized hopping parameter.\cite{Goth_2014} Note that this statement remains true in the presence of interactions since the transformation leaves the density $n_{j,\sigma} = c_{j,\sigma}^{\dagger} c_{j,\sigma} = d_{j,\sigma}^{\dagger} d_{j,\sigma} $ invariant and thus does not change the form of the interaction term $\mathcal{H}_{\text{int}}$. For repulsive electron-electron interaction, the system therefore would be in the Luttinger liquid phase (see Sec.~\ref{subsec:Renormalization group analysis}), as it would be in the absence of SOC.  

This possibility of gauging out the Rashba term was presented recently by Goth and Assaad in Ref.~\onlinecite{Goth_2014}.  However, it is curious to note that this Hamiltonian is identical to that of two spinless fermionic chains in a magnetic field.\cite{Narozhny_2005,Carr_Narozhny_Nersesyan_2006}  The Rashba term in the former model is equivalent to the orbital magnetic field in the latter; the inter-chain coupling in the ladder model being equivalent to a Zeeman term in the Rashba case, were such a term to be present.

In the ladder realization of the Hamiltonian, it is fairly clear why the magnetic field can be gauged out unless the chains are coupled: a pure one-dimensional object can have no orbital motion, and therefore can not feel the orbital effects of a magnetic field.  In its essence, the reason why spin-orbit may be gauged out in the Hubbard chain is analogous: There is no "orbital" motion possible, so the spin-orbit may only couple as a pure gauge [one should be careful, however, that this is not to say there are no observable effects of spin-orbit (see Ref.~\onlinecite{Goth_2014} for the analysis)]. 

However, as any physical manifestation of a one-dimensional wire is necessarily due to a confining potential in the transverse directions, the Rashba spin-orbit interaction may nevertheless couple to the wave functions in these transverse directions to affect the system in a thermodynamic manner.  This was analyzed 15 years ago in a series of papers by Moroz \textit{et al};\cite{Moroz_Barnes_1999,Moroz_2000_PRB,Moroz_2000_PRL} the crux of this work is that non-trivial effects occur when the parity (inversion) symmetry of the wire is broken in an essential way (assuming time-reversal symmetry is preserved).  This broken inversion symmetry may be intrinsic to the wire itself, or due to the confining potentials.  While we refer to these original papers for an overview of the materials physics aspect of this, it turns out one may capture this behavior with a simple toy model, \eqref{latticemodel}, with next-nearest neighbor hopping.

\begin{figure}
      \begin{center}
      \includegraphics[width=0.5\textwidth]{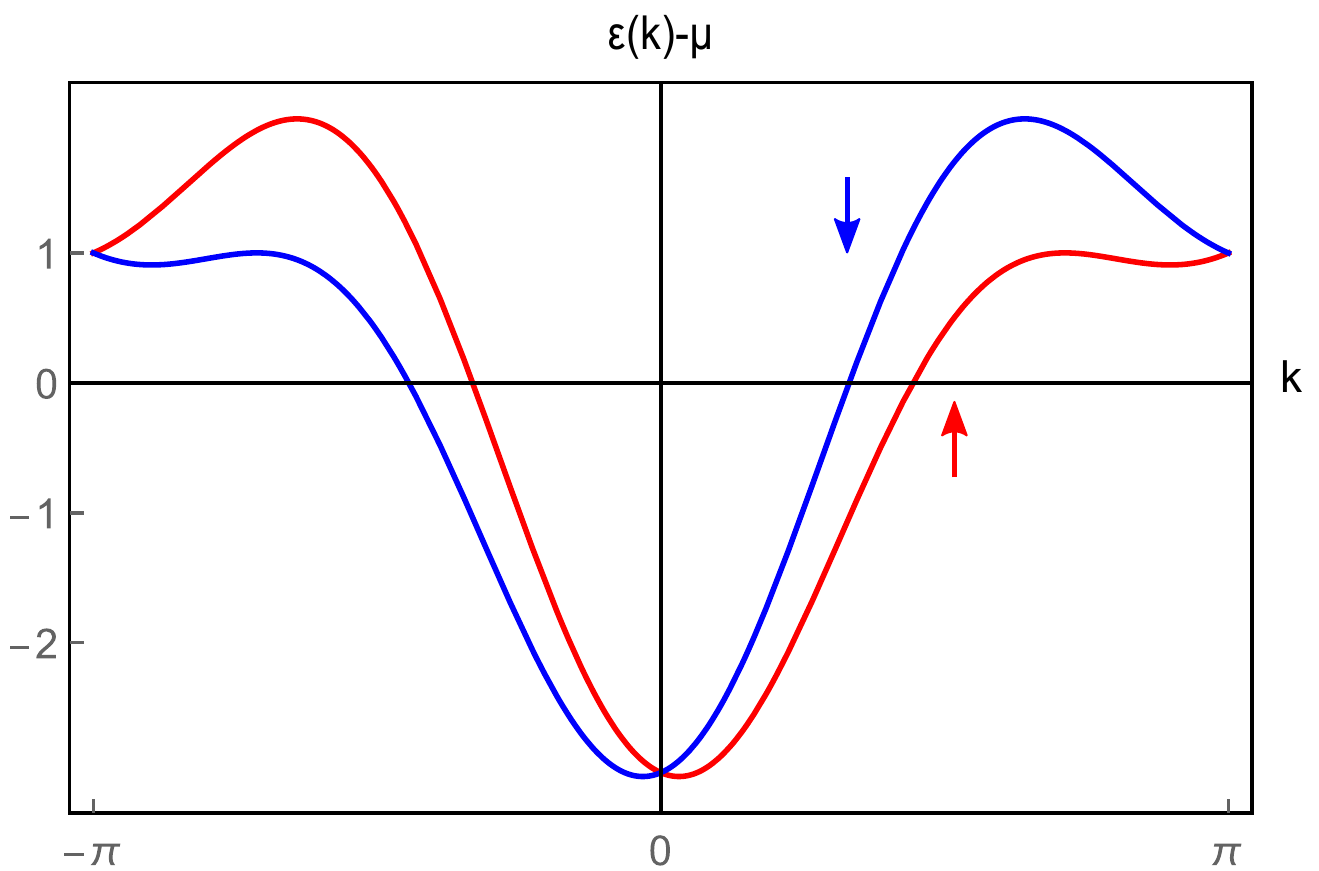}
         \caption{\small Spectrum in Eq.~(\ref{spectrum}) of the hopping model in Eq.~(\ref{latticemodel}) in the noninteracting case. The bands are labeled by the $z$-component of the spin and shifted by the constant momentum $q_0$. The energy in units of the hopping parameter $t$ is shown for the parameters $\alpha/t = 0.3 $ and $t'/t = 0.5$. 
                         \label{Fig:spectrum}}
      \end{center}   
\end{figure}

When $t'\neq 0$, the spectrum of \eqref{latticemodel} is given by
\begin{align}
   \epsilon_{\sigma}(k) = -2 \tilde{t} \cos(k- \sigma q_0) -2 t' \cos(2 k) \, . \label{spectrum}
\end{align}
We find two bands characterized by the $z$-component of the spin that are shifted by an constant momentum $q_0$. The band structure is depicted in Fig.~\ref{Fig:spectrum}. Notice that the inversion symmetry is broken due to the SOC $\epsilon_{\sigma}(k) \neq \epsilon_{\sigma}(-k)$. While this is also true when $t'=0$, in the presence of next-nearest neighbor hopping this leads to different Fermi velocities at the Fermi points of each band [cf. Fig. 2], and thus the symmetry can not be restored by a trivial gauge transformation. Meanwhile, time-reversal symmetry of the model is still preserved and therefore $\epsilon_{\sigma}(k) = \epsilon_{-\sigma}(-k) $.

We can find an analytical estimate for the velocity difference in the limit when $\tilde{t} \gg t'$ and the chemical potential is tuned to the bottom of the band. In this case, we can expand the spectrum around $k = \sigma q_0$. We find the Fermi points of the band $\sigma$ determined by the equation $\epsilon_{\sigma}(k_{1,2}) = \mu$ and the Fermi velocities:
\begin{align}
\begin{split}
    v_{1,2} =& \left. \frac{\partial \epsilon_{\sigma}(k)}{\partial k} \right|_{k = k_{1,2}} \\
              =&  \pm 2 \sqrt{(t')^2 \sin^2(2 q_0)+ \tilde{\mu} \left(\tilde{t} +t' \cos(2 q_0) \right) } \\
               & + 4 t' \sigma \sin(2 q_0) \, . \end{split}
\end{align}
where $\tilde{\mu} = 2 t' \cos(2 q_0)+2  \tilde{t}+\mu$. 
Due to the preserved time-reversal symmetry, the Fermi points and Fermi velocities are not independent but rather $k_{\uparrow,1} = - k_{\downarrow,1} \equiv k_1$, etc.
The velocity difference at the Fermi level is given by
\begin{align}
   \Delta v =  \frac{\abs{v_{1}} - \abs{v_{2}} }{\abs{v_{1}} + \abs{v_{2}}}  =  \frac{ 2 t'}{\tilde{t}} \sin(2 q_0) + \mathcal{O}\left[ \left( t'/\tilde{t}\right)^2 \right] \, .
\end{align}
Notice that the velocity difference vanishes either in the absence of SOC, $\alpha = 0$, or without next-nearest-neighbor hopping, $t' = 0$. For weak nearest-neighbor hopping and weak SOC, it is of the order of $\delta v \sim  \alpha t'/ t^2$.

To summarize, by using an explicit hopping model for one-dimensional fermions in the presence of SOC, we have identified two main effects of SOC on the spectrum of noninteracting electrons. First, it breaks the SU(2) spin-rotational symmetry and therefore lifts the spin degeneracy of the spectrum and second it breaks inversion symmetry leading to different Fermi velocities $v_1 \neq v_2$. Before we discuss the effect of interaction on the phase diagram of the system we now review existing work on the topic to put this work in the correct context.

\subsection{Summary of previous work}
 
This work considers two interrelated questions: first, whether spin-orbit coupling may drive a [otherwise SU(2) invariant] single-channel quantum wire with repulsive interactions to a spin-gapped SDW phase; and secondly, what are the physical properties of this phase, in particular, those associated with non-trivial topology.  As, particularly, the former of these questions has a rich history, we find it useful to briefly summarize previous results relevant to this work.



  The model we use was first described fifteen years ago by Moroz \textit{et al.}\citep{Moroz_Barnes_1999,Moroz_2000_PRB,Moroz_2000_PRL}, who derived explicitly that in general the SOC gives rise to two bands with different Fermi velocities.  They modeled a quantum wire as a two-dimensional electron gas confined in one spatial direction by an external potential and derived the band structure of the effective one-dimensional model. The spectrum in the presence of SOC turns out to have the same form as that of our hopping model and they propose the effective low-energy theory we discuss in Sec.~\ref{sec:Interaction effects}.  We note, however, that it was not until very recently\cite{Goth_2014} that it was realized how essential this difference in Fermi velocities is, as if this effect is neglected, the SOC may be removed by a gauge transformation.
  
  The effect of interactions was also discussed in the early works.\citep{Moroz_2000_PRB}  By a simple scaling dimension analysis, they concluded that the backscattering interaction term potentially responsible for the opening of a spin-gap is always irrelevant (for repulsive interactions), and thus a spin-gap never opens.  However, it turns out that in the vicinity of the SU(2) symmetric point, the scaling dimension alone is not a good indication of relevance as the backscattering interaction at this point is exactly marginal.\cite{GNT_book}  One should therefore study this question more carefully.
  
Five years later, Gritsev \textit{et al.}\citep{Gritsev_2005} revisited the problem of the RG of interacting fermions in the presence of Rashba SOC.  By treating the problem within two-loop RG, they concluded that the SDW phase is possible, and constructed a phase diagram.  However, they start from general parameters in the model [meaning interactions may explicitly break SU(2) symmetry even in the absence of spin-orbit coupling], and it is very difficult from their work to determine the line in parameter space the physical situation where the SU(2) is broken only by spin-orbit.  In fact, as they do not consider different velocities in the two bands, it must be true within their model that the spin-gap state is never realized on this line.

A few years later, Schulz \textit{et al.} revisited this problem within one-loop RG, but treating the marginal backscattering interaction much more carefully.\cite{Schulz_2009}  Through this analysis, they obtained a Berezinskii-Kosterlitz-Thouless (BKT) phase diagram (as is the case without spin orbit\cite{GNT_book}), but with renormalized effective parameters, which depend crucially on the velocity difference.  Importantly, if the velocity difference goes to zero, the bare backscattering parameters are not renormalized; which is an equivalent way of saying that the SOC may be gauged out.  The result of this calculation, however, was that the parameters are always renormalized towards the weak-coupling side of the phase diagram, implying once more that the SDW state is never realized.

We follow a very similar approach to Ref.~\onlinecite{Schulz_2009}, but by integrating out the gapless charge sector, we treat exactly the forward scattering couplings between the spin and charge sectors before applying perturbative RG.  Like Schulz \textit{et al.}, we also obtain a BKT phase diagram with renormalized parameters, but these renormalized parameters can now be on either side of the phase boundary, meaning that under certain conditions (which are derived in this work), the spin-gap may appear and the SDW phase is realized.


We note also that there is a lot of work on the realisation of the SDW state when a magnetic field is also added to the system.
The authors in Refs.~\onlinecite{Sun_2007,Gangadharaiah_2008} considered the situation when both spin-orbit coupling and a magnetic field are present which breaks SU(2) symmetry completely. They find that the system undergoes a phase transition to the SDW state if the magnetic-field and spin-orbit axes are orthogonal. 
Furthermore, if the developed spin gap is large enough, the ordering in the spin sector can crucially suppress the backscattering of electrons of nonmagnetic impurities. On the other hand, magnetic impurities can localize the SDW phase and destroy the (near-)perfect conduction properties. This is a general property of the SDW phase and will discussed in detail in our work in Sec.\ref{subsec:Effect of disorder}. One key feature of the present case is that this phase is robustly protected against these magnetic impurities by the time-reversal symmetry in the system, or in other words, as long as time-reversal symmetry remains unbroken, this state will remain metallic.


Finally, we comment briefly on the topological properties of the phase we find, which are similar to those of a topological insulator (due to the spin-gap), but occur in a metallic system, due to the gapless charge modes.  Recently, another work appeared by Keselman and Berg\cite{Keselman_2015} looking at exactly this question.  Although Keselman and Berg looked at attractive interactions, they were interested in a spin-triplet pairing rather than the conventional spin-singlet pairing, and this spin-triplet pairing has exactly the same spin structure as our SDW.  The topological properties of the two models are therefore almost identical. The fractional spin-edge states we find are analogous to those in \onlinecite{Keselman_2015}, while we believe our conduction properties will also carry over to their model.


\section{Interaction effects}
\label{sec:Interaction effects}
\begin{figure}
      \begin{center}
      \includegraphics[width=0.5\textwidth]{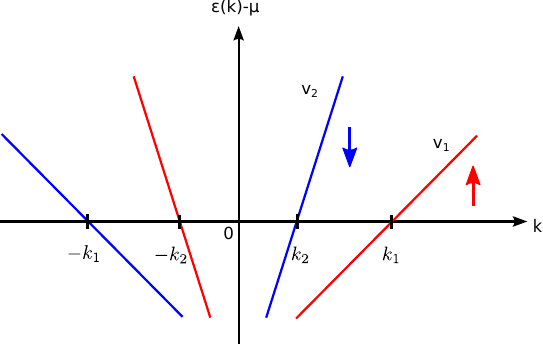}
         \caption{\small Band structure $\epsilon(k)$ of the low-energy theory. Due to time-reversal invariance the dispersions of the two bands are connected as
          $\epsilon_{\uparrow}(k) = \epsilon_{\downarrow}(-k)$. The low-energy excitations for up spins are right moving particles with velocity $v_1$ 
          at Fermi momentum $k_1$ and left moving particles with velocity $-v_2$ at Fermi momentum $-k_2$ (analogously for down spins).   
                         \label{Fig:lowenergyspectrum}}
      \end{center}   
\end{figure}

We now construct an effective low-energy Hamiltonian based on the previous results on SOC effects. In the previous section we have learned that the SOC lifts the spin degeneracy of the energy bands and leads to two bands ($\sigma = \uparrow, \downarrow$) with asymmetric Fermi velocities $ v_{1} \neq v_2$. 
To find the effective low-energy form of the Hamiltonian we linearize the spectrum near the Fermi points $k_{1,2}$, see ~Fig. \ref{Fig:lowenergyspectrum}. Next, we expand fermionic operators in modes that vary slowly on the scale of the inverse Fermi momentum: 
\begin{align}
\begin{split} 
   c_{j,\uparrow} \enspace \to \enspace \psi_{\uparrow}(x) =  R_{\uparrow}(x) \, e^{i k_{1} x}+ L_{\uparrow}(x) \, e^{-i k_{2} x} \, , \\
   c_{j,\downarrow} \enspace \to \enspace \psi_{\downarrow}(x) = R_{\downarrow}(x) \, e^{i k_{2} x}+ L_{\downarrow}(x) \, e^{-i k_{1} x}\, .
   \label{lowenergydecomposition}
\end{split}   
\end{align}
This yields the density
\begin{align}
   n_{j,\sigma} \to  R_{\sigma}^{\dagger} R_{\sigma} + L_{\sigma}^{\dagger} L_{\sigma}
                      +R_{\sigma}^{\dagger} L_{\sigma} e^{-i 2 k_F x}
                      +L_{\sigma}^{\dagger} R_{\sigma} e^{i 2 k_F x} \, , \label{density}
\end{align}
where we defined $ k_F = (k_1 + k_2)/2$. While the SOC breaks the chiral symmetry of the low- energy model, time-reversal is still preserved.
For spinful fermions, the time-reversal symmetry operation $\mathcal{T}$ in real space can be represented as $\mathcal{T} = i \sigma_y \mathcal{K}$. Here, $\sigma_y$ acts in spin space and $\mathcal{K}$ denotes complex conjugation. This implies $\psi_{\uparrow}(x) \to \psi^{\ast}_{\downarrow}(x)$ and $\psi_{\downarrow}(x) \to -\psi^{\ast}_{\uparrow}(x)$. Using the low-energy decomposition in Eq.~(\ref{lowenergydecomposition}) we find
$ R_{\uparrow}(x) \to L_{\downarrow}^{\ast}(x)$, $L_{\uparrow}(x) \to R_{\downarrow}^{\ast}(x)$ $R_{\downarrow}(x) \to -L_{\uparrow}^{\ast}(x) $, and 
$ L_{\downarrow}(x) \to -R_{\uparrow}^{\ast}(x)$. Using the bosonization conventions outlined in Appendix~\ref{App:Bosonization conventions}, this implies the following transformation properties in the spin-charge basis:
\begin{align}
\begin{split} 
  &\varphi_{c}(x) \; \to \; \varphi_{c}(x) \, , \quad \varphi_{s}(x) \; \to \; -\varphi_{s}(x)\, , \\
  &\theta_{c}(x) \; \to \; -\theta_{c}(x) \, , \quad \theta_{s}(x) \; \to \; \theta_{s}(x)\, .\end{split} \label{TRrules}
\end{align}
Additionally the Klein factors transforms as $\kappa_{\uparrow} \to \kappa_{\downarrow}$ and $\kappa_{\downarrow} \to -\kappa_{\uparrow}$.

The low-energy form of the Hamiltonian is given by
\begin{align}
\begin{split}   
  \mathcal{H}_0 =&  -i v_1 \int \! \mathrm{d} x \,  \left( R_{\uparrow}^{\dagger} \partial_x R_{\uparrow}^{}
                                               - L^{\dagger}_{\downarrow} \partial_x L_{\downarrow} \right) \\
       &  -i v_2 \int \! \mathrm{d} x \,  \left( R_{\downarrow}^{\dagger} \partial_x R_{\downarrow}
                                               - L^{\dagger}_{\uparrow} \partial_x L_{\uparrow} \right) \, , \end{split}  \\ 
\begin{split} 
  \mathcal{H}_{\text{int}} =& \, U  \int \! \mathrm{d} x \, \big(R_{\uparrow}^{\dagger} R_{\uparrow}^{} +L_{\uparrow}^{\dagger} L_{\uparrow}^{}  \big)
                            \big(R_{\downarrow}^{\dagger} R_{\downarrow}^{} +L_{\downarrow}^{\dagger} L_{\downarrow}^{}  \big) \\
                             +&U \int \! \mathrm{d} x \,         \big(R_{\uparrow}^{\dagger}
                             L_{\uparrow}^{} L_{\downarrow}^{\dagger} R_{\downarrow}^{} +\text{H.c.} \big)\, .                                                                                                  
  \label{model}
\end{split}   
\end{align}
Here, $v_1$ and $v_2$ are considered as phenomenological parameters of the low-energy theory which describe the different Fermi velocities at the left and right Fermi point. On the basis of our analysis of the hopping model in Sec.~\ref{sec:Tight binding model for one dimensional electrons in the presence of spin-orbit-coupling} we expect the velocity difference to be small but in principle tunable through the Rashba parameter $\alpha$.
The interaction term $\mathcal{H}_{\text{int}}$ follows from applying the decomposition~(\ref{density}) in the Hubbard interaction in Eq.~(\ref{HHubbard}).

\subsection{Bosonization}
\label{subsec:Bosonization}

We study the low-energy theory with the help of the bosonization technique using the conventions outlined in Appendix~\ref{App:Bosonization conventions}.
The resulting Hamiltonian density of the kinetic part reads as 
\begin{align}
\begin{split} 
   \mathcal{H}_0 =& \enspace  \frac{v_F}{2} \sum_{\sigma} \Bigl[ \left(\partial_x \varphi_{\sigma}\right)^2
                 + \Pi_{\sigma}^2 \Bigr] \\
                 &  +\frac{\delta v}{2} \Bigl[ \partial_x \varphi_{\uparrow} \Pi_{\uparrow} - \partial_x \varphi_{\downarrow}
                  \Pi_{\downarrow}\Bigr] \, ,\end{split}
\end{align}
where we introduced the difference $\delta v = v_1-v_2$ and average $v_F = (v_1+v_2)/2$ of Fermi velocities .

We now introduce the conventional spin and charge operators 
\begin{align}
   \varphi_c  = \frac{\varphi_{\uparrow} + \varphi_{\downarrow}}{\sqrt{2}}, \qquad 
   \varphi_s  = \frac{\varphi_{\uparrow} - \varphi_{\downarrow}}{\sqrt{2}}\, .
\end{align}
In the new basis the Hamiltonian density including electron-electron interaction reads as
\begin{align}
\begin{split}   
   \mathcal{H}_c =& \frac{v_c}{2} \Bigl[ K_c \Pi_{c}^2 + K_c^{-1}\left(\partial_x \varphi_{c}\right)^2
                   \Bigr]\, ,\\
   \mathcal{H}_s =& \frac{v_s}{2}  \Bigl[ K_s \Pi_{s}^2 +K_s^{-1}\left(\partial_x \varphi_{c}\right)^2
                    \Bigr] \\
                  & +\frac{g_s}{2 (\pi a)^2} \cos\left( \sqrt{8 \pi} \varphi_s \right)  \, ,         \\   
   \mathcal{H}_{\text{SO}} =& \frac{\delta v}{2} \Bigl[ \partial_x \varphi_c \Pi_s  + \partial_{x} \varphi_s \Pi_c
                       \Bigr] \, .
\label{Hamiltonian}                          
\end{split}                                                        
\end{align}
Here, the Luttinger parameters are defined as 
\begin{align}
   K_{\mu} =  1 + \frac{g_{\mu}}{2 \pi v_F} \, , \qquad \mu = c,s \, .
\end{align}
Due to Galilean invariance of the model it must hold that 
\begin{align}
   v_{\mu} K_{\mu} = v_F \, . \label{spinvelocity}
\end{align}
In terms of the original parameters, we have
\begin{align}
   g_c = -g_s = - a_0 U \, ,
\end{align}
where we reinstated the lattice constant $a_0$. In particular for repulsive interaction $U>0$, we find $K_c<1$.

We notice that the charge sector is described by a LL state with coupling constants $v_c$ and $K_c$.
The spin sector is also a LL but includes a backscattering term that can generate a spin gap if it becomes relevant in the RG sense. 
We point out that the different Fermi velocities and thus the SOC manifests itself only in the term $\mathcal{H}_{\text{SO}}$ that breaks the spin-charge separation.
In the absence of SOC the RG flow of the model is described by the well-known BKT equations (\ref{RGflow}) and the corresponding flow is constrained by SU(2) symmetry to the separatrix(see Fig.~\ref{Fig:BKT}).
Now, as long as the velocity asymmetry $\delta v$ is sufficiently small we can stay in the spin-charge basis and treat the term $\mathcal{H}$ as a small perturbation. Usually, the (marginal) term $\mathcal{H}_{SO}$ would be neglected as it only produces small corrections under the RG. In the present case we have to keep it since the conventional flow is exactly along the separatrix and even a marginal term may drive the system into a new phase. 

To study the effect of this perturbation on the RG flow of the model we will integrate out the (quadratic) charge sector
and derive the effective action of spin fields.

First we perform the Legendre transformation from the Hamiltonian to the Lagrangian
\begin{align}
\begin{split} 
   &\mathcal{L}\left[\varphi_i, \partial_x \varphi_i, \partial_t \varphi_i \right] +
   \mathcal{H}\left[\varphi_i, \partial_x \varphi_i, \Pi_i \right]\\
    & \qquad = \int \! \mathrm{d} x \, 
   \sum_i \Pi_i(x) \partial_t \varphi_i(x) \, . \label{Legendretransform}\end{split}
\end{align}
where $\partial_t \varphi_i = \delta \mathcal{H} / \delta \Pi_i$. Solving the equations for the conjugate momentum $\Pi$ yields the expressions
\begin{align}
   \Pi_c =& \frac{1}{v_F} \partial_{t} \varphi_c -\frac{\delta v}{2 v_F} \partial_x \varphi_s , \\
   \Pi_s =& \frac{1}{v_F} \partial_{t} \varphi_s -\frac{\delta v}{2 v_F} \partial_x \varphi_c \, .
   \label{conjugatemomentum}
\end{align}
Substituting Eqs.~(\ref{Hamiltonian}) and~(\ref{conjugatemomentum}) into Eq.~(\ref{Legendretransform}) yields the resulting Lagrangian density  
\begin{align}
\begin{split}    
   \mathcal{L}_c =& \frac{1}{2 v_c K_c} \left[ \left(\partial_t \varphi_c\right)^2  - 
                   v_c^2\left( 1- \frac{\delta v^2}{4 v_c^2} \right)\left(\partial_x \varphi_c \right)^2
                   \right] \; \\
                 =& \; \frac{1}{2 v^{\ast}_c K^{\ast}_c} \left[ \left(\partial_t \varphi_c\right)^2  - 
                   (v_c^{\ast})^2  \left(\partial_x \varphi_c \right)^2   \right] \, ,\\
   \mathcal{L}_s =& \frac{1}{2 v_s K_s} \left[ \left(\partial_t \varphi_s\right)^2  - 
                   v_s^2\left( 1- \frac{\delta v^2}{4 v_s^2} \right)\left(\partial_x \varphi_s \right)^2
                   \right]  \\
                  & - \frac{g_s}{2 (\pi a)^2} \cos\left( \sqrt{8 \pi} \varphi_s \right) \, ,\\             
   \mathcal{L}_{sc} =& -\frac{\delta v}{2 v_F} \left[ \partial_x \varphi_c \partial_t \varphi_s +
                       \partial_x \varphi_s \partial_t \varphi_c   \right]   \, .
\end{split}                                     
\end{align}
Here we introduced the renormalized coupling constants
\begin{align}
   K_c^{\ast} = K_c \left(1- \frac{\delta v^2}{4 v_c^2} \right)^{-\frac{1}{2}}, \qquad
   v^{\ast}_c = v_F/ K_c^{\ast}\, .
   \label{chargevelocity}
\end{align}
Next we determine the action in imaginary time
\begin{align}
\begin{split} 
   i S =& i \int \! \mathrm{d} t \mathrm{d} x \, \mathcal{L}\left[\varphi, \partial_x \varphi, 
       \partial_t \varphi \right] \enspace
     \\ \stackrel{t = i \tau}{\to}    \enspace
   -S =&  \int \! \mathrm{d} \tau \mathrm{d} x \, \mathcal{L}\left[\varphi, \partial_x \varphi, 
       i \partial_{\tau} \varphi \right] \, .
\end{split}       
\end{align}
The partition function is then
\begin{align}
   Z   =& \int \! \mathrm{D} \varphi_c \mathrm{D} \varphi_s \, e^{-S_s[\varphi_s]-S_c[\varphi_c]-S_{sc}
          [\varphi_c,\varphi_s]} \, , \\
   S_c \, =& \,  \int \! \frac{\mathrm{d} x \, \mathrm{d} \tau}{2 v_F} \; \left[
          (\partial_{\tau} \varphi_c)^2 + (v^{\ast}_c)^2 \, (\partial_x \varphi_c)^2 \right] \, , \\
   \begin{split}       
   S_s \, =& \,  \int \!  \frac{\mathrm{d} x \, \mathrm{d} \tau}{2 v_F} \, \big[
          (\partial_{\tau} \varphi_s)^2 +  \big\lbrace v_s^2- \frac{(\delta v)^2}{4 } \big\rbrace \, (\partial_x \varphi_s)^2 \big]         \\   
         &+ \frac{g_s}{2(\pi a)^2} \int \! \mathrm{d} x \mathrm{d} \tau  \, \cos\left( \sqrt{8 \pi}
          \varphi_s \right) \, , \end{split} \\ 
   S_{sc} \, =& \, \frac{i \delta v}{2 v_F} \, \int \! \mathrm{d} x \mathrm{d} \tau \,
             \left[ \partial_x \varphi_c \partial_{\tau} \varphi_s + \partial_x \varphi_s 
             \partial_{\tau} \varphi_c \right]  \, .  
\end{align}
Finally, integrating out the charge sector yields the effective action $S_s = S_0 +S_{\text{int}}$ in the spin sector at zero temperature : 
\begin{align}
\begin{split}  
   S_0 =& \frac{1}{2} \int \! \frac{\mathrm{d} q}{2\pi} \frac{\mathrm{d} \omega}{2\pi} \, \varphi_s(q,\omega) \varphi_s(-q,-\omega) \\
         &\left[ \frac{1}{v_s(q,\omega) K_s(q,\omega)} \omega^2 + \frac{v_s(q,\omega)}{K_s(q,\omega)} 
          q^2 \right]\, , \\  \\
   S_{\text{int}} =&  \frac{g_s}{2 (\pi a)^2} \int \! \mathrm{d} x \,\mathrm{d} \tau \;
                      \cos\left( \sqrt{8 \pi} \varphi_s \right)   \,.   
   \label{effectiveactionspin}
\end{split}    
\end{align}
The effective spin velocity and the Luttinger parameter obey the following equations
\begin{align}
   & v_s(q,\omega) K_s(q,\omega) = v_F \, , \\
   \frac{v_s(q,\omega)}{K_s(q,\omega)} =& \frac{v_s}{K_s} - v_F \left(\frac{\delta v}{2 v_F}\right)^2 
   \left[ 1 - \frac{4 \omega^2}{\omega^2 + (v_c^{\ast} q )^2} \right]\, . \label{rem1}
\end{align}
Note that $v_s$ and $K_s$ are the coupling constants of the system without SOC. The effective propagator in the spin sector is affected in two ways by SOC as can be seen in Eq.~(\ref{rem1}):

(i) The parameters in the spin sector are explicitly renormalized by a $\delta v$ term and (ii) there is a contribution from the charge degrees of freedom where $v_c^{\ast}$ is renormalized according to Eq.~(\ref{chargevelocity}).

In the following, we will derive the RG equations of the effective action Eq.~(\ref{effectiveactionspin}) employing a standard Wilson RG procedure in momentum space in order to study how SOC affects the phase diagram of interacting electrons.

\subsection{Renormalization group analysis}
\label{subsec:Renormalization group analysis}
\begin{figure}
      \begin{center}
      \includegraphics[width=0.37\textwidth]{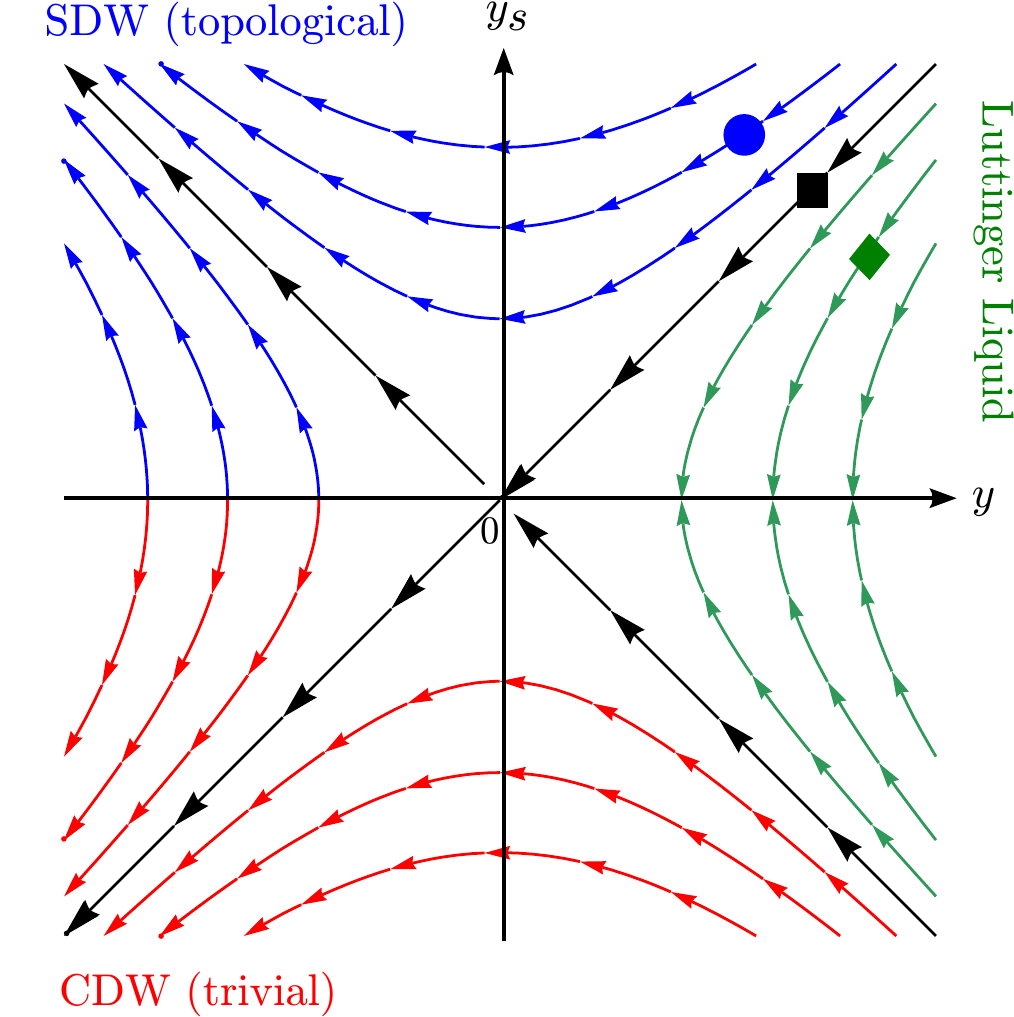}
         \caption{\small RG flow of Eq.~(\ref{RGflow}) describing the phase diagram of interacting 1D fermions. In the presence of spin-orbit coupling, the starting point of the flow is changed away from the SU(2) invariant line (black rectangle) to the region of strong coupling flow (blue circle) for strong spin-orbit coupling, $\overline{\delta v} > 2 \overline{U}$ or to the region of flow towards the Luttinger liquid phase for $\overline{\delta v} < 2 \overline{U}$ (green diamond). 
                         \label{Fig:BKT}}
      \end{center}   
\end{figure}

To determine the effect of SOC in the presence of interactions, we study the RG equations of the effective action Eq.~(\ref{effectiveactionspin}) employing a perturbative Wilson RG procedure in momentum space. To this end, we integrate out all high energy degrees of freedom between the momentum cutoff $\Lambda$ and a lower cutoff $\Lambda'$ to obtain the low-energy physics of the model. Details of the calculation can be found in Appendix \ref{App:Details of the momentum space RG}. The result of this procedure is encoded in differential equations for the dimensionless strength of backscattering $y_s$ and forward scattering $y$ (the latter is related to $K_s$). The equations are of the BKT type and read as:
\begin{align}
\begin{split}
   \frac{d y_s}{d\ell} \enspace =& -y_s(\ell) y(\ell)   \, ,\\ 
  \frac{d y}{d\ell} \enspace =& \enspace - y_s^2(\ell) \label{RGflow}    \, . \end{split}
\end{align}
Here, $\ell = \ln(\Lambda /\Lambda')$ and the initial values are given by
\begin{align}
  y_s(0) = \overline{U} \, \quad \text{and} \quad y(0) = \overline{U} + \overline{\delta v} \, f(\overline{U},\overline{\delta v}) \, . \label{initialvalues}
\end{align} 
They are determined by the dimensionless strength of interaction $\overline{U} = a_0 U / \pi v_F$ and the dimensionless strength of SOC $\overline{ \delta v} = (\delta v /2 v_F)^2$.
The function $f(\overline{U},\overline{\delta v}) $ is nonuniversal in the sense that it depends on the way the integration procedure of the RG is performed. The universal fact, however, is that it vanishes at some point and that 
\begin{align}
   \text{sign}\big( f(\gamma) \big) = \text{sign}\left( \frac{2 \overline{U}}{\overline{\delta v}}-1\right) \, .
\end{align}
The ratio of the two dimensionless parameters $\overline{U}$ and $\overline{ \delta v}$ therefore completely determines the phase of the interacting electron gas in the presence of SOC. The corresponding phase diagram is shown in Fig.~\ref{Fig:BKT}. In the absence of spin-orbit coupling, for $\overline{\delta v} =0$, the bare coupling constants are equal, $y_s(0)=y(0)$, and the flow is along the separatrix. The SOC changes the initial values of the coupling strengths and depending on the sign of the function $f(\overline{U},\overline{\delta v})$ we  either have $y_s(0)>y(0)$ or $y_s(0)<y(0)$. For strong SOC, $\overline{\delta v} > 2 \overline{U}$, the sign of $f$ is negative, which results in $y_s(0)>y(0)$ and the system will flow to the strong coupling SDW phase. On the other hand if $\overline{\delta v} < 2 \overline{U}$ we have $y_s(0)<y(0)$ and the system flows to a Luttinger liquid phase with renormalized Luttinger parameters.

To show this, we now review some of the properties of the BKT flow equations.\cite{GNT_book}
The flow equations are invariant under a sign change of $y_{s}$ (but not $y$) and are characterized by the invariant $\mu^2 = y_{s}^2 -y^2$ and the ratio of initial coupling constants $y(0)/y_{s}(0) \equiv \cos(\beta)$. We point out that while the RG flow is independent of the sign of $y_s$ the nature of the strong coupling fixed point is not. Indeed as we will discuss later in Sec.~\ref{subsec:The nature of the strong coupling phase} the dominant correlations in the strong coupling phase are either of the spin-density wave type ($y_s < 0$) or the charge-density wave type ($y_s >0$).
  
The flow equations can be integrated to find
\begin{align}
   y(l) = \mu \cot(\mu l + \beta), \qquad y_{s}(l) = \frac{\mu}{\sin(\mu l + \beta)} \, .  \label{solutionBKT}
\end{align}
There are three different regimes, illustrated in Fig.~\ref{Fig:BKT}.
\begin{enumerate}

\item[(I)] \textbf{weak coupling}: $|y(0)| > y_{s}(0), y(0)> 0$. Here $\beta = i \chi$, $\chi >0$ and $\mu = i m$ where $m>0$. Backscattering flows to weak coupling, $y_{s} \to 0$, while $y \to m$ as $l \to \infty$. This phase is realized for $\overline{\delta v} < 2 \overline{U}$.

\item[(II)] \textbf{cross-over} Here $\mu >0$ and $0 \leq \beta \leq \pi$. The flow is still to strong coupling but via an intermediate regime ($0 \leq \beta \leq \pi/2$) where $y_{s}$ initially decreases.  This phase is realized for $\overline{\delta v} > 2 \overline{U}$.

\item[(III)] \textbf{strong coupling}: $|y(0)| > y_{s}(0), y(0)< 0$. Here $\beta = i \chi + \pi$, $\chi >0$ which yields $\mu = -i m$ where $m>0$. Both coupling constants flow to strong coupling reaching a pole singularity at $l_0 = \chi/m$. This phase is not realized in the context of our weak coupling analysis.

\end{enumerate}

The lines where $\mu = 0$ are the SU(2) symmetric lines. The flow is to weak coupling for repulsive interaction [$y(0) >0$] and towards strong coupling for attractive interaction [$y(0) <0$]. The RG flow evolves along these lines in the absence of SOC.

We are particularly interested in the study of the strong coupling phase where a gap opens in the spin sector of the theory. This phase is realized in the case of moderately strong SOC, $\overline{\delta v} > 2  \overline{U}$, i.e., the crossover regime in the above classification. 

To get an estimate of the magnitude of the spin-gap $\Delta_s$, we integrate the RG-flow Eq.~(\ref{solutionBKT}) up to a scale $l^{\ast}= \ln(\Lambda/\Delta_s)$ where $g_s(l^{\ast}) = 1$. In our case, the deviation from the separatrix is always small and thus $\mu \simeq \overline{\delta v}^2 \ll 1$. This yields the estimate
\begin{align}
   \Delta_s = \Lambda e^{- \beta/ \mu} \, ,
\end{align}
where $0 <\beta < \pi$. The spin gap is therefore exponentially small.

\subsection{Nature of the strong coupling phase}
\label{subsec:The nature of the strong coupling phase}

In this section, we discuss which quasi-long-range correlations govern the behavior of the strong coupling phase.
According to the Mermin-Wagner theorem, there is no spontaneous breaking of a continuous symmetry and thus no long-range order in dimensions $D \leq 2$ at any finite temperature and for short range interactions. Therefore, the average of any order parameter $\braket{\mathcal{O}(x)}$ vanishes.
On the other hand, an operator $\mathcal{O}$ may exhibit quasi-long-range order meaning that the correlation function $\braket{\mathcal{O}(x) \mathcal{O}(0)}$ decays as a power law. The ground state of the system is then characterized by the operator whose correlations decays the slowest among all.

We study two possibilities for the order parameter:
\begin{enumerate}
   \item Charge-density wave:
   \begin{align}
   \begin{split}
         \mathcal{O}_{\text{CDW}} =& \Big( R_{\uparrow}^{\dagger} L_{\uparrow}^{} +L_{\uparrow}^{\dagger} R_{\uparrow}^{} \Big) + 
                              \Big( R_{\downarrow}^{\dagger} L_{\downarrow}^{} +L_{\downarrow}^{\dagger} R_{\downarrow}^{} 
                              \Big)\\
                           =& -\frac{1}{\pi a} \big[  \sin(\sqrt{4 \pi} \varphi_{\uparrow}) + \sin(\sqrt{4 \pi}
                               \varphi_{\downarrow}) \big]\\
                           =& -\frac{2}{\pi a} \sin(\sqrt{2 \pi} \varphi_{c}) \cos(\sqrt{2 \pi} \varphi_{s}) \, . 
                           \label{CDW} 
   \end{split}                        
   \end{align}
   \item $z$-component of the spin-density wave:
   \begin{align}
   \begin{split}
        \mathcal{O}_{\text{SDW}} =& \Big( R_{\uparrow}^{\dagger} L_{\uparrow}^{} +L_{\uparrow}^{\dagger} R_{\uparrow}^{} \Big) - 
                              \Big( R_{\downarrow}^{\dagger} L_{\downarrow}^{} +L_{\downarrow}^{\dagger} R_{\downarrow}^{} \Big)\\
                           =& -\frac{1}{\pi a} \big[  \sin(\sqrt{4 \pi} \varphi_{\uparrow}) - \sin(\sqrt{4 \pi} \varphi_{\downarrow}) \big]\\
                           =& -\frac{2}{\pi a} \cos(\sqrt{2 \pi} \varphi_{c}) \sin(\sqrt{2 \pi} \varphi_{s}) \, . 
                           \label{SDW} 
   \end{split}                        
   \end{align}
\end{enumerate}   
Superconducting correlations are also possible, but we will not consider them here since we discuss the experimentally relevant case of repulsive electron-electron interaction $K_c<1$ for which their correlations decay faster than density wave correlations.  We note, however, that the crucial element in the following discussion is the structure in spin-space (which has two possibilities if gapped), so it is easy to adapt our results for the case of attractive interactions, for example, as discussed in Ref.~\onlinecite{Keselman_2015}.

In the strong coupling phase, the field $\varphi_s$ orders and therefore its dual field $\theta_s$ is totally disordered and all its correlation functions decay exponentially to zero (this is the case for the $x$ and $y$ components of the SDW and we therefore exclude them from our discussion). On the other hand, the CDW and SDW order parameters will develop quasi-long range order. 

We are now in the position to discuss the nature of the strong coupling phase of interacting electrons in the presence of SOC.
We note that if the cosine term in the spin-sector is relevant (massive phase), we have two possibilities:
\begin{itemize}
   \item $g_s <0$, in which case the field $\varphi_s$ wants to lock at a minimum where $\cos(\sqrt{8 \pi} \varphi_s) = 1$ to minimize the energy of the system,
         i.e. $\varphi_s = \sqrt{\pi/2} n$. In this case $\braket{\cos(\sqrt{2 \pi } \varphi_s)} \neq  0$ and  $\braket{\sin(\sqrt{2 \pi } \varphi_s)} =  0$, so
         the dominant correlations are of the CDW type. 
   \item $g_s >0$, in which case the field $\varphi_s$ wants to lock at a minimum where $\cos(\sqrt{8 \pi} \varphi_s) = -1$ to minimize the energy of the system,
         i.e. $\varphi_s = \sqrt{\pi/2} (n+1/2)$. In this case $\braket{\cos(\sqrt{2 \pi } \varphi_s)} =  0$ and  $\braket{\sin(\sqrt{2 \pi } \varphi_s)} \neq
         0$, so the dominant correlations are of the SDW type.       
\end{itemize}

In the case of repulsive electron-electron interaction discussed above, we have $g_s(0) \simeq \overline{U} > 0$. The system therefore develops SDW correlations in the $z$-component of the spin.

\section{Properties of the spin-density-wave phase}
\label{sec:Properties of the spin-density wave phase}

\subsection{Effect of disorder}
\label{subsec:Effect of disorder}

Consider the situation that the system has established spin-density-wave order, i.e., the spin sector of the theory is gapped. We now want to study the effect of sparsely distributed disorder in this phase. This situation can be modeled by considering a local scattering potential at $x=0$ with strength $\lambda_{\sigma}$ for up- and down electrons, respectively:
\begin{align}
\begin{split} 
   H_{\text{imp}} =& \sum_{\sigma} \lambda_{\sigma} n_{\sigma}(0) \\
                  =& \sum_{\sigma} \lambda_{\sigma} \big[ (R_{\sigma}^{\dagger} R^{}_{\sigma} + L_{\sigma}^{\dagger} 
                     L^{}_{\sigma})+( R_{\sigma}^{\dagger} L^{}_{\sigma} +\text{H.c.}) \big] \, .
\end{split}                     
\end{align}
The first term describes forward scattering and we will neglect it from now on since it is well known that it can be removed via a gauge transformation.\cite{GNT_book} Thus,
\begin{align}
\begin{split} 
   H_{\text{imp}}  =& \sum_{\sigma} \lambda_{\sigma} ( R_{\sigma}^{\dagger} L^{}_{\sigma} +\text{H.c.}) \\
                   =& - \frac{1}{\pi a} \big[ \lambda_{\uparrow} \sin(\sqrt{4 \pi} \varphi_{\uparrow}) + \lambda_{\downarrow} 
                      \sin(\sqrt{4 \pi} \varphi_{\downarrow})  \big]\\
                   =& - \frac{\lambda_{\uparrow } + \lambda_{\downarrow}}{ \pi a} \sin(\sqrt{2 \pi} \varphi_{c}) 
                   \cos(\sqrt{2 \pi} \varphi_{s})   \\
                    &  - \frac{\lambda_{\uparrow } - \lambda_{\downarrow}}{ \pi a} 
                   \sin(\sqrt{2 \pi} \varphi_{s}) \cos(\sqrt{2 \pi} \varphi_{c}) \\
                   =& \enspace  \frac{\lambda_{\uparrow } + \lambda_{\downarrow}}{ 2} \mathcal{O}_{\text{CDW}} +
                      \frac{\lambda_{\uparrow } - \lambda_{\downarrow}}{ 2} \mathcal{O}_{\text{SDW}}  \, .
\end{split}                      
\end{align}
In the last line we expressed the impurity operator in terms of the CDW and SDW order parameters defined in Eqs.~(\ref{CDW}) and~(\ref{SDW}).
We point out that according to the transformation rules in Eq.~(\ref{TRrules}) $\mathcal{O}_{\text{CDW}}$ is even under time-reversal while $\mathcal{O}_{\text{SDW}}$ is odd (note that this does not imply that time-reversal symmetry is spontaneously broken in the SDW phase, since the type of order is not long-range).
We therefore refer to the two terms as nonmagnetic and magnetic impurity, respectively. Time-reversal symmetry of the Hamiltonian constricts the disorder potential to be symmetric in spins, $\lambda_{\uparrow} = \lambda_{\downarrow} = \lambda$. Therefore, only the nonmagnetic part of the impurity remains and the
impurity operator directly couples to the CDW order parameter operator:
\begin{align}
    H_{\text{imp}}  = \enspace  \lambda \mathcal{O}_{\text{CDW}}  \, .
\end{align}
Since charge-density-wave correlations vanish in the SDW phase, we can already anticipate that impurities will be an irrelevant perturbation to the system. This statement can be made more precise by the following argument. 
In the spin-gap phase, the field $\varphi_{s}$ is locked to one of the minima of $\cos(\sqrt{8 \pi} \varphi_s)$. Fluctuations around this ground state can be described semiclassically by expanding
\begin{align}
    \varphi_s(x,\tau) = \varphi_s^{(0)} + \delta \varphi_s(x,\tau)  \, .
\end{align}
In the case of repulsive electron-electron interaction discussed in this work, the minima are $\varphi_s^{(0)} = \sqrt{\pi/2} (n+1/2)$ and thus we get
\begin{align}
    H_{\text{imp}}  =& \enspace  \frac{\sqrt{2} \lambda}{\sqrt{\pi} a} \sin(\sqrt{2 \pi} \varphi_{c}) \delta \varphi_s + \mathcal{O}(\delta \varphi_s^3) \, .
\end{align}
Integrating out the massive fluctuations $\delta \varphi_s$ the charge sector of the model maps to the Kane-Fisher model\cite{Kane_Fisher_1992} with $K \to 2 K_c$. Thus nonmagnetic impurities have no effect to first order but can generate a term in second order that can become relevant for very repulsive interactions $K_c < 1/2$. Magnetic impurities, on the other hand, which will only be present when time-reversal symmetry is broken, are given by 
\begin{align}
   H_{\text{imp,magn}} =  \frac{\lambda}{ \pi a}  \cos(\sqrt{2 \pi} \varphi_{c}) + \mathcal{O}(\delta \varphi_s^2)  \, ,
\end{align}
which becomes relevant for $K_c<2$. So even a small concentration of magnetic impurities would destroy the SDW state. 

The analysis of a single impurity in the SDW phase closely resembles the study of a local perturbation in a fermionic ladder.\citep{Carr_Narozhny_Nersesyan_2011, Carr_Narozhny_Nersesyan_2013} In the latter case, the system consists of two interacting and closely located nanowires with CDW correlations which have a relative phase shift of $\pi$. The authors find the surprising result that a single impurity placed on a rung of the ladder represents an irrelevant perturbation that does not change the conductance at zero temperature. On the other hand, two identical impurities placed on the same rung represent a relevant perturbation driving the ladder to an insulating phase.  This is completely analogous to the nonmagnetic or magnetic impurity in this work.

Let us now discuss the situation when we are not dealing with single impurities but the impurity density is high. The corresponding Hamiltonian is given by
\begin{align}
\begin{split}
   H_{\text{imp}} =& \sum_{\sigma} \int \! \mathrm{d} x \, U(x) n_{\sigma}(x) \\
                  =& \sum_{\sigma} \int \! \mathrm{d} x \, \Big[ U_f(x) (R_{\sigma}^{\dagger} R_{\sigma} + L_{\sigma}^{\dagger} L_{\sigma}) \\   
                   & \hspace{1cm} + ( U_b(x) R_{\sigma}^{\dagger} L{\sigma} + \text{H.c.})\Big]  \\
                  =& \frac{2}{\sqrt{\pi}} \int \! \mathrm{d} x \, \partial_x \varphi_c U_f(x) \\
                   & - \frac{i}{4 \pi} \int \! \mathrm{d} x \, U_b(x) 
                     e^{- \sqrt{2 \pi} \varphi_c} \cos(\sqrt{2 \pi} \varphi_s) +\text{H.c.} \, . \end{split}
\end{align}
Here we defined the forward and backward scattering potentials
\begin{align}
\begin{split}
   U_f(x) =& \int_{|q| \ll k_F} \! \frac{\mathrm{d} q}{2 \pi} \, e^{i q x} U(q)\, , \\
   U_b(x) =& \int_{|q| \ll k_F} \! \frac{\mathrm{d} q}{2 \pi} \, e^{i q x} U(q+2 k_F) \, .  \end{split}
\end{align}
which are white noise correlated $\overline{U_f(x) U_f(x')} = \overline{U_b(x) U^{\ast}_b(x')} =  D \delta(x-x') $. After averaging over disorder the complete model is
\begin{align}
\begin{split} 
   S_c =& \frac{1}{2 K_c^{\ast} v_c^{\ast}} \sum_a \int \! \mathrm{d} x \mathrm{d} \tau \, \big[ (\partial_{\tau} \varphi_c^a)^2 + (v_c^{\ast})^2 
          (\partial_{x} \varphi_c^a)^2 \big]  \, , \\
   S_s =& \frac{1}{2 K_s^{\ast} v_s^{\ast}} \sum_a \int \! \mathrm{d} x \mathrm{d} \tau \, \big[ (\partial_{\tau} \varphi_s^a)^2 + (v_s^{\ast})^2    (\partial_{x} \varphi_s^a)^2 \big]  \\
        & +  \frac{g_s}{2 (\pi a)^2} \sum_a\int \! \mathrm{d} x \,\mathrm{d} \tau \;
                      \cos\left( \sqrt{8 \pi} \varphi_s^a \right)   \,, \\ 
   S_{SO} =& i \frac{\delta v}{v_F} \sum_a\int \! \mathrm{d} x \mathrm{d} \tau  \,  \partial_{\tau} \varphi_c^a \partial_x \varphi_s^a    \, ,   \\
   S_f =& -\frac{2}{\sqrt{\pi}}  \sum_a \int \! \mathrm{d} x \mathrm{d} \tau  \, U_f(x) \partial_x \varphi_c^a \, , \\
   S_b =& \frac{D}{(\pi a)^2}   \sum_{a,b} \int \! \mathrm{d} x \mathrm{d} \tau_1 \mathrm{d} \tau_2  \, \cos(\sqrt{2 \pi} [\varphi_c^a(1)-\varphi_c^b(2)]) \\
        & \enspace \times     \cos(\sqrt{2 \pi} \varphi_s^a(1))  \cos(\sqrt{2 \pi} \varphi_s^b(2))   \, .   
\end{split}     
\end{align}
Here, $a$, $b$ denote replica indices.
First we note that forward scattering can be removed by the gauge transformation $\varphi_c^a \to \varphi_c^a + K_c^{\ast}/ v_c^{\ast} \sqrt{\pi} \int^x \! \mathrm{d} y \, U_f(y) $. Second, we consider the case where the spin gap already has established and we add disorder on top of it. In this case we may neglect the marginal term $S_{SO}$ and expand $\varphi_s$ around the ground state configuration. Integrating out the massive fluctuations $\delta \varphi_s$ the model maps to the Giamarchi-Schulz model\cite{Giamarchi_Schulz_1988} with $K \to2 K_c^{\ast}$. Disorder thus becomes relevant for $K_c^{\ast} < 3/4$.

This result can also be implied from the result for a single impurity by the following heuristic reasoning.
For a single nonmagnetic impurity the conductance reads as
\begin{align}
   G = \frac{2 e^2}{h} - A \epsilon^{4 K_c -2}  \, ,
\end{align} 
where $A$ is a nonuniversal constant and $\epsilon = \max[T,V]$. To establish the boundary between localized and delocalized regime for the case of weak disorder, it is sufficient to replace $\epsilon \to 1/L$ and multiply $\delta G$ by the number of impurities $N_{imp} \sim 1/L$. Now, if G grows with increasing $L$, the system is in the localized regime or else it is in the delocalized regime. This yields the thresshold $K_c < 3/4$ for localization in the presence of nonmagnetic impurities and $K< 3$ for magnetic impurities. This situation is very similar to the effect of disorder in quantum-spin-Hall edge states.\cite{Kainaris_2014} The edge states are also protected against nonmagnetic impurities for weak interactions, $K_c>3/8$, due to the time-reversal symmetry but are expected to localize if the interactions in the system become too strong. 

\subsection{Luther Emery solution and edge states}
In the bosonized theory it is possible to investigate the properties of the strong coupling phase by refermionization (see Appendix~\ref{App:Bosonization conventions}) which enables us to map the sine Gordon model at $K_s= 1/2$ to spinless fermions with mass $m = g_s / 2 \pi  a$. The Hamiltonian density in the spin sector at $K_s=1/2$ then reads as
\begin{align}
\begin{split} 
   \mathcal{H}_{\text{s}} =& \frac{v}{2} \big[ \Pi^2 + (\partial_x \varphi)^2 \big] + \frac{m}{\pi a} \cos(\sqrt{4 \pi}
                              \varphi)  \\
                           =& -i v \big( R^{\dagger}\partial_x R^{}  -  L^{\dagger}\partial_x L^{} \big)  + i m \big(
                               R^{\dagger}  L^{} - L^{\dagger} R^{}  \big) \label{LutherEmery1}
\end{split}                               
\end{align}
While this is valid only at one specific value of interaction strength, we will assume that the features of the refermionized Hamiltonian characterize the whole massive sector of the model. One particular property of the model Eq.~(\ref{LutherEmery1}) we want to mention at this point is that it hosts zero energy modes localized at the edge when put on a finite segment $0 \leq x \leq L$ with open boundary conditions.\cite{Fabrizio_1995}
In the $L \to \infty$ limit (semi-infinite system), the wave function of the zero energy bound state at $x=0$ takes the form
\begin{align}
    \chi_{0}(x) = \frac{m}{v_s} e^{-\frac{m}{v_s} |x|} 
\end{align}
While it is not mentioned explicitly in the original publication, it is important to realize that the edge state is only a normalizable solution if $m >0$. 

Since electrons carry both charge and spin, the edge state in the spin sector affects the whole electron liquid .
To see this, consider a boundary between a topologically trivial phase with $m>0$ (e.g., the vacuum) and the topologically nontrivial phase with $m<0$ at $x=0$. Since the field $\varphi_s$ is pinned to $\varphi_s = \sqrt{\pi/2} n_1$ for $m<0$ and to $\varphi_s = \sqrt{\pi/2} (n_2+1/2)$ for $m>0$ where $n_1$, $n_2$ are integers, there must be a kink of minimal magnitude $\sqrt{\pi/8}$ in $\varphi_s$ across the boundary. Such a kink in $\varphi_s$ corresponds to an accumulation of half of the electron spin at the boundary:
\begin{align}
\begin{split} 
   S_z =& \int \! \mathrm{d} x \, \rho_s(x) = \frac{1}{\sqrt{2 \pi}} \int \! \mathrm{d} x \, \partial_x \varphi_s(x) \\
       =&   \frac{1}{\sqrt{2 \pi}} \left[ \varphi_s(0-) - \varphi_s(0+)\right] =   \pm \frac{1}{4} \, .
\end{split}       
\end{align} 
where we used the spin-density defined in Eq.~(\ref{spindensity}).  This agrees with recent findings in an analogous model in Ref.~\onlinecite{Keselman_2015}.

\subsection{Topological classification} 
\label{sec:Topological classification}

In this section, we show that the the strong coupling fixed point in the spin sector of interacting electrons in the presence of SOC is a topological insulator of class BDI.

The topological phase of a model of noninteracting massive fermions can be determined by the transformation properties of its single-particle Hamiltonian, which is defined in the first Brillouin zone, under the anti-unitary operations of time-reversal $\theta$, charge conjugation $\mathcal{C}$, and the combined chiral operation $\Xi \sim \theta \mathcal{C}$. They are defined as 
\begin{align}
\begin{split}
   \theta \mathcal{H}^{\ast}(k) \theta^{-1} =& +\mathcal{H}(-k) \, , \qquad \theta^2 = \pm 1 \, \\
   \mathcal{C} \mathcal{H}^{\ast}(k) \mathcal{C}^{-1} =& - \mathcal{H}(-k) \, , \qquad \mathcal{C}^2 = \pm 1 \, \\
    \Xi \mathcal{H}^{\ast}(k) \Xi^{-1} =&- \mathcal{H}(k) \, . \label{antiunitarysymmetries} \end{split}
\end{align}
Consider now the refermionized Hamiltonian~(\ref{LutherEmery1}), which describes the effective theory of the spin sector in the massive phase. Defining the vector $\Psi = (R, L)^{T}$ the Hamiltonian in momentum space takes the form
\begin{align}
      H =  \Psi^{\dagger}(k) \big( v k \sigma_z - m \sigma_y \big) \Psi(k) =  \Psi^{\dagger}(k) \mathcal{H}(k)\Psi(k) \, . \label{invariantlowenergymodel}
\end{align}  
For the single particle Hamiltonian $\mathcal{H}(k)$ in Eq.~(\ref{invariantlowenergymodel}), we can explicitly construct
the antiunitary operations in Eq.~(\ref{antiunitarysymmetries}). They can be represented as $ \theta = \sigma_x \mathcal{K}$, $\mathcal{C} = \mathcal{K}$, $\Xi = \sigma_x$, where $\mathcal{K}$ denotes complex conjugation and the Pauli matrices act in chiral space. Note, that these are symmetries of the refermionized spin sector, in particular the operator $\theta$ represents the projection of the time-reversal operator onto the gapped spin sector and should not be confused with the time-reversal operator $\mathcal{T}$ of the physical electrons defined above Eq.~(\ref{TRrules}). From the explicit construction of the operators, we see that $\theta^2 = \Pi^2 = \mathcal{C}^2 = 1$, which places the system into the topological class BDI. This class has a $\mathbb{Z}$ topological invariant in one dimension.

We now proceed to calculate the topological invariant. To this end, we bring the Hamiltonian in the ``conventional" chiral form by switching $\sigma_z$ to $\sigma_x$ by an appropriate unitary rotation. The resulting Hamiltonian reads as
\begin{align}
   \mathcal{H}'(k) = \begin{pmatrix} 0 & h(k) \\ h^{\ast}(k) & 0\end{pmatrix} = \bs{d}(k) \bs{\sigma}                    
\end{align}
where we defined $h(k) = v k + i m$ and $\bs{d}(k) = (v k , -m,0)^T$. We define the normalized matrix element of the Hamiltonian $q(k) = h(k) / |\bs{d(k)}|$ and the winding number
\begin{align}
\begin{split}
   \nu =& \;  \frac{i}{2 \pi} \int_{-\infty}^{\infty} \! \mathrm{d} k \, q^{-1}(k) \partial_k q(k)
       =\;   \frac{1}{2} \text{sign}(m) \, .
\end{split}       
\end{align}
The winding number is not an integer since it is originally defined as a mapping from the compactified Brillouin zone, isomorphic to $\mathcal{S}^1$, to $\mathbb{Z}$. The low-energy theory defined on a non compactified manifold ($\mathbb{R}$) does not capture the physics in the whole Brillouin zone but only at one time-reversal-invariant point [the other half of the winding number is contributed by the other time-reversal-invariant point but we have no means to tell if it contributes $\text{sign}(m)$ or $-\text{sign}(m)$]. From the winding number we can therefore not deduce which sign of the mass corresponds to the topological and which to the trivial phase. We can, however, determine the number of edge states at the boundary of two topological materials with masses $m_1$ and $m_2$. It is given by $n = \nu(m_1)-\nu(m_2)$ and since $\nu \neq 0$, we know that one sector has to be topologically nontrivial. In fact, we have already established that the SDW phase with $m <0$ hosts zero energy edge states and therefore this has to be the topologically nontrivial one while the CDW phase is a topologically trivial insulating phase.
Note that the $\mathbb{Z}$ topological invariant can take only two values in the present case because we consider only a single quantum wire which can either have edge states or not. If we considered multiple identical wires, scattering between them cannot gap the individual edge states as long as the anti-unitary symmetries are preserved leading to the $\mathbb{Z}$ topological classification. This is similar to the topological properties of multiple copies of the Su-Schrieffer-Heeger model\cite{SSH_1979} for polyacetylene. We want to stress that the above classification scheme applies only to the gapped spin sector, which can be mapped to noninteracting fermions. Nonetheless, it allows us to clearly identify that a topological and a trivial phase exist. The properties exhibited by the whole electron liquid if the spin sector is in the topological phase have been discussed in the previous sections.

\section{Summary} 
\label{sec:Summary}
We investigated the influence of spin-orbit coupling in interacting one-dimensional quantum wires.
Spin-orbital interactions break the spin-rotational symmetry from SU(2) to U(1) and break the inversion symmetry in the wire.
This lifts the spin degeneracy of the spin-up and -down energy bands and leads to different Fermi velocities $v_1\neq v_2$ in each band which is depicted in Fig.\ref{Fig:lowenergyspectrum}. 
In a specific hopping model discussed in Sec.~\ref{sec:Tight binding model for one dimensional electrons in the presence of spin-orbit-coupling}, we estimate that the velocity difference is of the order of $\delta v \simeq \alpha t' /t^2$, where $\alpha$ is the strength of SOC and $t$, $t'$ are the amplitudes of nearest- and next-nearest neighbor hoppings, respectively. We find that this difference in velocities drastically affects the nature of interacting electrons.

The most interesting situation occurs when the dimensionless strength of SOC is stronger than the strength of interactions $\overline{\delta v} > 2 \overline{U}$.
In this case, interactions drive the spin sector of the system to a strong coupling phase with quasi-long-range spin-density-wave order where a spectral gap is dynamically generated. This prediction was established by treating interactions using bosonization and a weak coupling renormalization group analysis which is controlled by the small parameters of dimensionless interaction $\overline{U}$ and spin-orbit-coupling strength $\overline{\delta v}$.

We show that the spin sector of the gapped spin-density-wave state is topologically nontrivial (symmetry class BDI). Meanwhile, the charge sector of the quantum wire is still massless.  Since physical electrons carry both spin and charge the whole electron liquid forms an unusual topological state. The topological nature manifests itself in the emergence of zero-energy edge modes at the boundary of a finite system that carries fractional electron spin. The existence of these modes is protected by the bulk spin gap. Furthermore, we find that due to the gap in the spin sector, the bulk of the system is protected against nonmagnetic impurity scattering for interaction strengths $K_c> 3/4$, for high impurity concentration and for $K_c>1/2$ in the case of a single impurity. These extraordinary transport properties are protected by the time-reversal symmetry in the system.

We hope that our work will stimulate experimental search and studies of strongly interacting one-dimensional systems with sizable spin-orbit coupling. In particular, the robustness against disorder and the gapless edge states of the system should be probable in experiment.  We also note that this robustness to disorder may eventually be technologically exploitable in nano-electronics, as the near-perfect conduction property of the wire is protected by topology and does not require ultra-clean wires.

\section{Acknowlegdgements} 
\label{sec:Acknowlegdgements}

We would like to thank I.~Gornyi, A.~Mirlin, M.~Scheurer, and P.~Strange for useful discussions and suggestions at various stages of this work. N.K. thanks the Carl-Zeiss-Stiftung for financial support. This work was supported by the program DFG SPP 1666 ``Topological Insulators".

\appendix

\section{Bosonization conventions}
\label{App:Bosonization conventions}

We use the following conventions:
\begin{align}
   R_{\sigma}(x) = \frac{\kappa_{\sigma}}{\sqrt{2 \pi a}} e^{i \sqrt{4 \pi } \phi^{R}_{\sigma}(x)}, \quad
   L_{\sigma}(x) = \frac{\kappa_{\sigma}}{\sqrt{2 \pi a}} e^{-i \sqrt{4 \pi } \phi^{L}_{\sigma}(x)}.
\end{align}
where $a$ is the bosonic UV cutoff. The Klein factors anticommute $\left\lbrace \kappa_{\sigma}, \kappa_{\sigma'} \right\rbrace = 2 \delta_{\sigma,\sigma'}$ and ensure correct anti-commutation relations of fermions with different spins. Since they are not dynamic variables, we are free to choose a specific representation:
\begin{align}
   \kappa_{\sigma}^2 = 1 , \qquad \kappa_{\uparrow} \kappa_{\downarrow} = - \kappa_{\downarrow} \kappa_{\uparrow} = i.
\end{align}
The correct anti-commutation relations of right- and left-movers are fixed by the commutations relations between $\phi^R$ and $\phi^L$:
\begin{align}
   [\phi^R_{\sigma},\phi^L_{\sigma'}] =&  \frac{i}{4}\delta_{\sigma, \sigma'} \, , \\
    [\phi^{\eta}_{\sigma}(x),\phi^{\eta'}_{\sigma'}(y)] =&  \frac{i}{4} \, \eta \, \delta_{\eta,\eta'} \delta_{\sigma, \sigma'} \sign(x-y) \, .
\end{align}
and the Campbell-Baker-Hausdorff rule 
\begin{align}
 e^{A} e^{B} = e^{A+B} e^{\frac{1}{2}[A,B]}
\end{align}
which holds when the commutator of $A$ and $B$ is not an operator. We also introduce the conjugate variables
\begin{align}
    \varphi_{\sigma} = \phi^R_{\sigma} + \phi^L_{\sigma}, \qquad 
   \theta_{\sigma} = \phi^L_{\sigma} - \phi^R_{\sigma} \, ,
\end{align}
where $\Pi = \partial_x \theta$ is the conjugate momentum to $\varphi$.
Using the above definitions, we find
\begin{align}
\begin{split} 
   \psi_{\eta,\sigma}^{\dagger} \partial_x \psi_{\eta,\sigma}^{} &= i \eta (\partial_x \phi_{\sigma}^{\eta})^2 \, ,\\
   R_{\sigma}^{\dagger} L_{\sigma} &= -\frac{i}{2 \pi a} e^{-i \sqrt{4 \pi} \varphi_{\sigma}} \, ,\\
   R^{\dagger}_{\sigma} R_{\sigma} +L^{\dagger}_{\sigma} L_{\sigma} &= 
   \frac{1}{\sqrt{\pi}} \partial_x \varphi_{\sigma} \, .
\end{split}      
\end{align}
The spin and charge degrees of freedom are given by 
\begin{align}
   \varphi_c  = \frac{\varphi_{\uparrow} + \varphi_{\downarrow}}{\sqrt{2}}, \qquad 
   \varphi_s  = \frac{\varphi_{\uparrow} - \varphi_{\downarrow}}{\sqrt{2}}\, .
\end{align}
In terms of these operators, the nonoscillatory part of the charge density and the $z$-component of the spin density read as
\begin{align}
\begin{split} 
   \rho_c(x) =& \sum_{\sigma} \psi_{\sigma}^{\dagger}(x) \psi_{\sigma}^{}(x)  =  \frac{\sqrt{2}}{\sqrt{\pi}} \partial_x \varphi_c(x) \, ,\\
   \rho^{z}_s(x) =& \sum_{\sigma}   \psi_{\sigma}^{\dagger}(x) S_{\sigma \sigma'}^z\psi_{\sigma'}^{}(x) =  \frac{1}{2}\sum_{\sigma}  \sigma \psi_{\sigma}^{\dagger}(x) \psi_{\sigma}^{}(x)  \\ =& \frac{1}{\sqrt{2 \pi}} \partial_x \varphi_s(x) \, .\label{spindensity}\end{split}
\end{align}
Finally, we state the following refermionization identities for spinless fermions:
\begin{align}
    e^{i \sqrt{4 \pi} \varphi} \enspace  \leftrightarrow& \enspace  -i 2 \pi a \, L^{\dagger} R^{} \, , \\
    e^{i \sqrt{4 \pi} \theta} \enspace  \leftrightarrow& \enspace  i 2 \pi a \, L^{\dagger} R^{\dagger} \, , \\
    \partial_x \varphi \enspace \leftrightarrow& \enspace \sqrt{\pi} \big( R^{\dagger} R^{} +L^{\dagger} L^{}  \big) \, , \\
    \Pi \enspace \leftrightarrow& \enspace \sqrt{\pi} \big( L^{\dagger} L^{} -R^{\dagger} R^{}   \big) \, , 
\end{align}

\section{Details of the momentum space RG}
\label{App:Details of the momentum space RG}

We introduce the vector notation $\textbf{q}=(q,\omega/v_{s})$ and $\textbf{r}=(x,v_{s} \tau)$. The absolute value is denoted by $r = |\textbf{r}|$.
To perform the RG analysis of Eq.~(\ref{effectiveactionspin}) we split the fields into fast ($>$) and slow ($<$) components (we drop the subscript $s$ from now on since we exclusively deal with the spin sector): 
\begin{align}
   \varphi(x,\tau) =& \enspace  \varphi^{>}(x,\tau) + \varphi^{<}(x,\tau)\, ,
\end{align}
where we defined
\begin{align}   
   \varphi^{<}(x,\tau) \enspace =& \enspace  \int\limits_{|\textbf{q}|<\Lambda/b} \hspace{-.2cm}
   \frac{\mathrm{d} q \mathrm{d} \omega}{(2 \pi)^2}  \enspace e^{i \textbf{q} \textbf{r}} \;
   \varphi(q,\omega) \, ,\\
   \varphi^{>}(x,\tau) =&  \int\limits_{\Lambda/b<|\textbf{q}|<\Lambda} \hspace{-.2cm}
   \frac{\mathrm{d} q \mathrm{d} \omega}{(2 \pi)^2}  \enspace e^{i \textbf{q} \textbf{r}} \;
   \varphi(q,\omega)\, .
\end{align}
Now, we integrate out the fast degrees of freedom whose momenta lie between the momentum cutoff $\Lambda$ and the new smaller cutoff $\Lambda/b$, defined by the scaling factor $b= e^{\ell} \approx 1+\ell >1$.  This yields the effective action 
\begin{align}
   S_{\text{eff}} =& S^{<}_{0} + \braket{S_{\text{int}}}_> - \frac{1}{2} 
                    \left[ \braket{(S_{\text{int}})^2}_> - \braket{S_{\text{int}}}_>^2
                    \right] \, , \\ \nonumber \\
   S_{\text{int}} =&  \frac{g_s}{2 (\pi a)^2} \int \! \mathrm{d} x \,\mathrm{d} \tau \;
                      \cos\big[ \sqrt{8 \pi} \left( \varphi^< + \varphi^> \right) \big] \, .    \label{SeffRG}            
\end{align}
To perform the average, we need the correlation function of spin fields in momentum and frequency space which can be obtained from Eq.~(\ref{effectiveactionspin}) and reads as
\begin{align}
\begin{split} 
   & \braket{\varphi(\bs{q}) \varphi(-\bs{q})} \, \\ =& \; \frac{K_s}{v_s} \frac{1}{(\omega/v_s)^2 +q^2 - q^2 \delta \left[1-\frac{4 (\omega/v_s)^2}{(\omega/v_s)^2+\gamma^2 q^2} \right]}\, .\end{split}
\end{align}
Here, we defined the dimensionless constants
\begin{align}
   \gamma =  \frac{v_{c}^{\ast}}{v_s} \, , \qquad \text{and} \qquad 
   \delta = \left( \frac{\delta v}{2 v_s}\right)^2 \, .
\end{align}
The velocities in the charge and spin sectors are defined in the main text in Eqs.~(\ref{spinvelocity}) and~(\ref{chargevelocity}).
They can be expressed in terms of microscopic parameters as
\begin{align}
   \overline{U} = \frac{a U}{\pi v_F} \, , \quad \text{and} \qquad \overline{\delta v} = \left(\frac{\delta v}{2 v_F}\right)^2 \, .
\end{align}
In order to make analytical progress, we assume weak interactions $\overline{U} \ll 1$ and weak SOC $\overline{\delta v} \ll 1$. This allows us to make the expansion
\begin{align}
   \gamma \simeq \, 1 + \overline{U} - \frac{1}{2} \overline{\delta v}  \, , \qquad \text{and} \qquad 
   \delta \simeq \,  \overline{\delta v} \, .
\end{align}
Furthermore, we define the functions:
\begin{align}
   F_{\gamma,\delta}(\Lambda r) \; =& \; \frac{2 \pi }{ \ell K_{s}}  \braket{\varphi^>(x,\tau) 
                               \varphi^>(0,0)}_> \, ,\\
   \begin{split}
   A_{\gamma,\delta}(\Lambda r) \; =& \; e^{-4 \pi \braket{\varphi^>(x,\tau)\varphi^>(0,0)}_>} \\
                      \; =& \;  e^{-2 \ell K_{s} F_{\gamma,\delta}(\Lambda r)}  \, .  \label{DefA}\end{split}
\end{align}

In general, both $F_{\gamma,\delta}(r)$ and $A_{\gamma,\delta}(r)$ will depend on the parameters $\gamma$ and $\delta$.
Before we proceed to derive the RG equations we shortly comment on the properties of the function $F_{\gamma,\delta}(r)$.
We first consider the case of $\delta = 0$. Then, we find $F(\Lambda r) =  J_0(\Lambda r)$.
The zeroth Bessel function of the first kind is oscillating and falls off as a power law as $r \to \infty$. This is, however, not the behavior we would like to have since it does not allow us to perform a gradient expansion in $r$. It was shown that this is a consequence of the fact that we chose a hard cutoff for the radial integration. If we were instead to choose a smooth cutoff $F(r)$ is truly short range.\cite{GNT_book} In the following, we will not specify the explicit form of $F(r)$ but just assume that it falls off rapidly as $r \to \infty$. 

We now calculate the function $F_{\alpha,\delta}(0)$ which turns out to determine the flow in Eq.~(\ref{flowequations}):
\begin{widetext}
\begin{align}
\begin{split} 
     F_{\gamma,\delta}(0) 
  \; =& \; \frac{2 \pi }{\ell K_s} \int\limits_{\Lambda/b<|\textbf{q}|<\Lambda} \hspace{-.2cm}
           \frac{\mathrm{d} q \mathrm{d} \omega}{(2 \pi)^2}  \enspace 
           \frac{K_s}{v_s} \frac{1}{(\omega/v_s)^2 +q^2 - q^2 \delta \left[1-\frac{4 (\omega/v_s)^2}{
           (\omega/v_s)^2+\gamma^2 q^2} \right]} \\
  \; =& \; \frac{1 }{\ell } \int_{\Lambda/b}^{\Lambda} \frac{\mathrm{d} q}{q}
           \int_0^{2 \pi} \! \frac{\mathrm{d} \theta}{2 \pi} \,
           \enspace \frac{1}{1- \delta \cos^2(\theta) \left[1-\frac{4 \sin^2(\theta)}{
           \sin^2(\theta)+\gamma^2 \cos^2(\theta)} \right]}         \\
  \; =& \; \int_0^{2 \pi} \! \frac{\mathrm{d} \theta}{2 \pi} \,
           \enspace \frac{1}{1- \delta \cos^2(\theta) \left[1-\frac{4 }{
           1+\gamma^2 \cot^2(\theta)} \right]}\, .
\label{DefinitionF}           
\end{split}                                
\end{align}
\end{widetext}
\textbf{Analytical solution in the limit of small $\delta$} 

The integral in Eq.~(\ref{DefinitionF}) cannot be performed analytically but we can find a analytical expression in the case of $\delta \ll 1$. To first order in $\delta$ we find:
\begin{align}
\begin{split} 
  & F_{\gamma,\delta}(0) \\
  \; \simeq& \; \int_0^{2 \pi} \! \frac{\mathrm{d} \theta}{2 \pi} 
           \left\lbrace 1+\delta \cos^2(\theta) \left[1-\frac{4 }{
           1+\gamma^2 \cot^2(\theta)} \right] \right\rbrace   \\    
     =& \;  1+ \frac{\delta}{2} \left[1-\frac{4}{(1+\gamma)^2} \right] 
     \; = \; 1+ \delta f(\gamma)   \, . 
\label{ApproximationF}
\end{split}               
\end{align}
In the last equation we defined
\begin{align}
   f(\gamma) = \frac{1}{2}-\frac{2}{(1+\gamma)^2} \, .
\end{align}
Notice that $f(\gamma)> 0$ for $\gamma>1$, $f(\gamma)< 0$ for $\gamma<1$ and $f(1)=0$ (at this value of $\gamma$ the lowest non vanishing correction is of order $\delta^2$). This can be summarized as
\begin{align}
   \text{sign}\big[ f(\gamma) \big] = \text{sign}\left( \frac{2 \overline{U}}{\overline{\delta v}}-1\right)
\end{align}
\textbf{Derivation of the RG equations} 

We now proceed to derive the RG equations by calculating the effective action in Eq.~(\ref{SeffRG}). To first order we find
\begin{align}
\begin{split} 
   S_{\text{eff}}^{(1)} =&  \frac{g_s}{2 (\pi a)^2} \int \! \mathrm{d} x \,\mathrm{d} \tau \;
                            \braket{\cos\big[ \sqrt{8 \pi} \left( \varphi^{<} + \varphi^{>}\right) \big]}_>
                             \\
                        =&  \frac{g_s}{2 (\pi a)^2} \int \! \mathrm{d} x \,\mathrm{d} \tau \;
                            \cos( \sqrt{8 \pi} \varphi^<) \enspace  
                            e^{-4 \pi \braket{\left[ \varphi^>(x,\tau) \right]^2}_>}   \\
                        =&  \frac{g_s}{2 (\pi a)^2} (2-2 K_s F_{\gamma,\delta}(0)) \ell
                            \int \! \mathrm{d} x \,\mathrm{d} \tau \;
                            \cos( \sqrt{8 \pi} \varphi) \, .
\label{RGcorrection1}
\end{split}                                                                
\end{align}
In the last step we rescaled space time $(x, \tau) \to b (x,\tau)$, so that the theory is again defined with the cutoff $\Lambda$. We now define the short hand notation $1=(x_1,\tau_1)$. The second order is 
\begin{widetext}   
\begin{align}
\begin{split}    
   S_{\text{eff}}^{(2)} =& - \frac{1}{2} \frac{g^2_s}{4 (\pi a)^4} \int \! \mathrm{d} 1 \,\mathrm{d} 2 \;
                            \Big\lbrace
                            \braket{\cos\big[ \sqrt{8 \pi} \left( \varphi^{<} + \varphi^{>}\right)(1) \big]
                            \cos\big[ \sqrt{8 \pi} \left( \varphi^< + \varphi^>\right)(2) \big]}_> \\
                         & \hspace{3cm}   -\braket{\cos\big[ \sqrt{8 \pi} \left( \varphi^< 
                           + \varphi^> \right)(1) \big]}
                            _{>} \braket{ \cos\big[ \sqrt{8 \pi} \left( \varphi^{<} + \varphi^{>}\right)(2)
                             \big]}_>   \Big\rbrace \\
                        =& - \frac{1}{2} \frac{g^2_s}{(2\pi a)^4} \int \! \mathrm{d} 1 \,\mathrm{d} 2 \;
                           \sum_{\epsilon_1,\epsilon_2 = \pm} e^{i \sqrt{8 \pi} (\epsilon_1 \varphi^{<}(1) 
                           + \epsilon_2 \varphi^<(2))}
                           \Big\lbrace \braket{ e^{i \sqrt{8 \pi} [\epsilon_1 \varphi^{>}(1) 
                           +\epsilon_2 \varphi^{>}(2)]} }_> - e^{-8\pi \braket{[\varphi^>(0)]^2}_>} 
                           \Big\rbrace \\
                        =& - \frac{g^2_s}{(2\pi a)^4} A^{-2}(0)  \int \! \mathrm{d} 1 \,\mathrm{d} 2 \;
                           \Big\lbrace \cos\left( \sqrt{8 \pi} [\varphi^<(1)-\varphi^<(2)] \right) \left[
                           A^{-2}(\Lambda r_s) -1\right] \\
                         &  \hspace{3.8cm}+ \cos\left( \sqrt{8 \pi} [\varphi^<(1)+
                           \varphi^<(2)] \right) \left[A^{2}(\Lambda r_s) -1\right] \Big\rbrace  \, ,  
\end{split}                                                   
\end{align}
\end{widetext}
where the function $A(\Lambda r)$ is defined in Eq.~(\ref{DefA}). We now introduce relative and center coordinates
\begin{align}
   X=&\frac{x_1+x_2}{2}, \quad x=x_1-x_2, \\  T =& \frac{\tau_1+\tau_2}{2}, \quad \tau = \tau_1-\tau_2   
\end{align}
and the vectors $\textbf{R} = (X, v_s T)$ and $\textbf{r} = (x, v_s \tau)$. This gives
\begin{align}
\begin{split}   
   &S_{\text{eff}}^{(2)} = - \frac{g^2_s}{(2\pi a)^4 v_s^2} A^{-2}(0)  \int \! \mathrm{d}^2 R  \,
                             \mathrm{d}^2 r \; \\ &
                           \Big\lbrace \cos\left( \sqrt{8 \pi} [\varphi^<(\textbf{R}+\textbf{r}/2)
                           -\varphi^<(\textbf{R}-\textbf{r}/2)] \right) 
                           \left[ A^{-2}(\Lambda r) -1\right] \\
                         &  + \cos\left( \sqrt{8 \pi} [\varphi^<(\textbf{R}+\textbf{r}/2))+
                           \varphi^<(\textbf{R}-\textbf{r}/2)] \right) \left[A^{2}(\Lambda r) -1\right]
                           \Big\rbrace  \, .                    
\end{split}                            
\end{align} 
We expand the terms in brackets to first order in $\ell$
\begin{align}
      A^{-2}(0)\left[ A^{-2}(\Lambda r) -1\right]   
     \simeq&  \enspace 2 \ell   K_s F_{\alpha,\delta}(\Lambda r_s)\, , \\ 
      A^{-2}(0)\left[ A^{2}(\Lambda r_s) -1\right] \simeq& -2 \ell K_s F_{\gamma,\delta}(\Lambda r)    \, .                                               
\end{align}
Since $F(r)$ is decaying rapidly as $r \to \infty$ we can perform a gradient expansion of the cosine terms.
\begin{align}
\begin{split} 
   &\cos\left( \sqrt{8 \pi} [\varphi^<(\textbf{R}+\textbf{r}/2)-\varphi^<(\textbf{R}-\textbf{r}/2)] \right) 
   \\ & \quad \simeq \enspace  1- 4\pi \Big( \nabla_{\textbf{R}} \varphi(\textbf{R}) \textbf{r} \Big)^2 \, , \end{split}\\
\begin{split} 
   &\cos\left( \sqrt{8 \pi} [\varphi^<(\textbf{R}+\textbf{r}/2)+\varphi^<(\textbf{R}-\textbf{r}/2)] \right) 
   \\ & \quad \simeq \enspace \cos\left( \sqrt{32 \pi} \varphi^<(\textbf{R})\right)   \, .\end{split}   
\end{align}
The cosine term that appears is less relevant than the original cosine term and we therefore neglect it.
The first term however gives the following correction to the quadratic action
\begin{align}
\begin{split}    
   S_{\text{eff}}^{(2)} =& \frac{8 \pi g_s^2 K_s }{(2 \pi a)^4 v_s^2} \ell
                           \int \! \mathrm{d}^2 R \, \Big( \nabla_{\textbf{R}} \varphi(\textbf{R}) \Big)^2 
                           \int \! \mathrm{d}^2 r \enspace  r^2 F_{\gamma,\delta}(\Lambda r) \, .
\end{split}                        
\end{align}
The integral over relative coordinates gives a nonuniversal number $A_1$. We define
\begin{align}
   \int \! \mathrm{d}^2 r \enspace  r^2 F_{\gamma,\delta}(\Lambda r)
  = \frac{2 \pi}{\Lambda^4}   A_{1}(\gamma, \delta)  \, .
\end{align} 
Since the correction is already linear in $\ell$ we can set the rescaling factor of the coordinates $b$ to one. The resulting correction reads as
\begin{align}
\begin{split}    
   S_{\text{eff}}^{(2)} =& \frac{g_s^2 K_s }{(a \Lambda)^4  \pi^2 v_s} A_{1}(\gamma, \delta) \ell
                           \int \! \mathrm{d} x \mathrm{d} \tau \, \Big[ (\partial_x \varphi)^2
                           + \frac{1}{v_s^2} (\partial_{\tau} \varphi)^2 \Big]\, .
\label{RGcorrection2}                           
\end{split}                        
\end{align}
The term gives a correction to $v_s/K_s$ but leaves $1/v_s K_s$ invariant, which means that only $K_s$ will flow. From Eqs.~(\ref{RGcorrection1}) and~(\ref{RGcorrection2}) we find the differential equations
\begin{align}
   \frac{d K_s}{d\ell} \; =& \; -\frac{g_s^2(\ell) K_s^3(\ell)}{(a \Lambda)^4 v_s^2} A_1(\gamma,\delta) \, ,\\
   \frac{d g_s}{d\ell} \; =& \;  (2-2 K_s(\ell) F_{\gamma,\delta}(0)) g_s(\ell)\, .
\end{align} 
We can bring them in the conventional BKT form by defining the dimensionless coupling constant 
\begin{align}
   y_s^2  \; = \;  \frac{2 g_s^2 K_s}{(a \Lambda)^4 \pi^2 v_s^2} A_1(\gamma,\delta) \, .
\end{align}
It is straightforward to show that, to first order in $g_s$, the RG equations for $y_s$ are not affected under this redefinition. The total set of flow equations then reads as
\begin{align}
\begin{split} 
   \frac{d y_s(\ell)}{d\ell} \enspace =& \enspace \big[2-2 K_s(\ell)F_{\gamma,\delta}(0)\big] y_s(\ell) \, ,
   \\ 
  \frac{d K_s(\ell)}{d\ell} \enspace =& \enspace - \frac{y_s(\ell)^2 K_s(\ell)^2}{2}    \, .                           
\label{flowequations}  
\end{split}   
\end{align}   
\noindent
Notice that these equations are perturbative in $g_s$ but exact in $K_s$ and the parameters $\gamma = v_c^{\ast}/v_s$ and $\delta= (\delta v / 2v_s)^2$.

We simplify the flow equations for small interaction strength and small $\delta$ by setting $K_s(\ell) = 1+ g(\ell)/2$ and using the asymptotic form of $F_{\gamma,\delta}$ derived in Eq.~(\ref{ApproximationF}). This yields
\begin{align}
\begin{split} 
   \frac{d y_s}{d\ell} \enspace =& \enspace  -y_s(\ell) \Big( g(\ell) + \delta \,  f(\gamma) \Big) 
                                 + \mathcal{O}(y_s g \delta)    \, ,  \\ 
  \frac{d y}{d\ell} \enspace =& \enspace - y_s^2(\ell)      \, .                        
\end{split}      
\end{align}
We can define $y(\ell) = g(\ell) + \delta \,  f(\gamma)$ which leaves us with the conventional KT flow equations.
\begin{align}
   \frac{d y_s}{d\ell} \enspace =& -y_s(\ell) y(\ell)   \, ,\\ 
  \frac{d y}{d\ell} \enspace =& \enspace - y_s^2(\ell)     \, .
\end{align}
The initial values of these equations are given by $y_s(0) \equiv \overline{U}$ and $y(0) = \overline{U} + \overline{\delta v} \, f(\gamma)$. Note that in the absence of SOC ($\delta \to 0$) the RG equations have to coincide to describe the flow along the separatrix of the BKT flow. Therefore, we must demand that $g(0) = y_s(0)$.


\bibliography{database}

\end{document}